\documentclass[10pt,journal,compsoc]{IEEEtran}
\ifCLASSOPTIONcompsoc
  \usepackage[nocompress]{cite}
\else
  \usepackage{cite}
\fi
\usepackage{booktabs} 
\usepackage{xurl}
\usepackage{xspace}
\usepackage[font=normal]{caption}
\usepackage{subcaption}
\usepackage{float}
\usepackage{multirow,makecell}
\usepackage{pifont}
\usepackage{tikz}
\usepackage{bbm}
\usepackage[inline]{enumitem}
\newcommand{\rev}[1]{\textcolor{black}{#1}}
\usepackage{xcolor}
\usepackage{listings}
\usepackage{tikz}
\usetikzlibrary{tikzmark}
\usetikzmarklibrary{listings}
\usepackage{amsmath}
\usepackage{amssymb}
\usepackage{amsmath}
\usepackage{graphicx}
\usepackage{makecell}
\usepackage{colortbl}
\definecolor{darkgreen}{rgb}{0.0, 0.3, 0.13}
\definecolor{darkred}{rgb}{0.2, 0.0, 0.13}
\usepackage{tcolorbox}
\tcbset{textmarker/.style={%
        parbox=false,boxrule=0mm,boxsep=0mm,arc=0mm,
        outer arc=0mm,left=3mm,right=3mm,top=3pt,bottom=3pt,
        toptitle=1mm,bottomtitle=1mm}
        }
        
\newtcolorbox{blueBox}{textmarker,
    colback=blue!10!white}
\usepackage[english]{babel}
\usepackage{blindtext}
\usepackage{xspace}
\usepackage{multicol}

\lstdefinestyle{myStyle}{
  belowcaptionskip=1\baselineskip,
  breaklines=true,
  language=C++,
  showstringspaces=false,
  basicstyle=\footnotesize\ttfamily,
  keywordstyle=\bfseries\color{green!40!black},
  commentstyle=\itshape\color{purple!40!black},
  identifierstyle=\color{blue},
  stringstyle=\color{orange},
  numbers=left,
  firstnumber=1,
}

\usepackage{hyperref}
\hypersetup{
	plainpages=false,
	colorlinks,
	urlcolor=blue,
	linkcolor=blue,
	citecolor=blue,
	bookmarksnumbered
}
\usepackage{tikz}
\usepackage{tabularx}
\usepackage{colortbl}
\usepackage{soul,xcolor}
\sethlcolor{yellow}
\usepackage{orcidlink}

\usepackage{tikz}
\usetikzlibrary{arrows.meta,positioning,shapes}

\newcommand{\etal}{{\em et al.}\xspace}
\newcommand{\xl}[1]{\Xhline{#1\arrayrulewidth}}

\newcommand{\BfPara}[1]{{\noindent\bf #1.}\xspace}
\newcommand{\eg}{{\em e.g.},\xspace}
\newcommand {\ie}{{\em i.e.},\xspace}

\makeatletter
\@ifundefined{startTableTextColor}{%

}{}
\makeatother

\lstdefinestyle{redcode}{
    basicstyle=\ttfamily\color{black},
    breaklines=true,
    frame=none,
    showstringspaces=false
}

\begin{document}
\title{Security and Quality in LLM-Generated Code: A Multi-Language, Multi-Model Analysis}

\author{Mohammed~F.~Kharma\orcidlink{0000-0001-8280-3285
}, Soohyeon~Choi\orcidlink{0009-0002-1252-2263}, Mohammad~Alkhanafseh\orcidlink{0000-0002-6250-7291
}, David~Mohaisen\orcidlink{0000-0003-3227-2505}

\IEEEcompsocitemizethanks{\IEEEcompsocthanksitem David Mohaisen and Soohyeon Choi are with the Department of Computer Science, University of Central Florida, Orlando, FL 32816 USA.
E-mail: mohaisen@ucf.edu (Corresponding author), soohyeon.choi@ucf.edu. \IEEEcompsocthanksitem Mohammed F. Kharma and Mohammad Alkhanafseh are with the Department of Computer Science, Birzeit University, Ramallah, Palestine. E-mail: mkharmah@birzeit.edu, malkhanafseh@birzeit.edu.}

\thanks{Manuscript received April 19, 2005; revised August 26, 2015.}}

\markboth{IEEE Transactions on Dependable and Secure Computing,~Vol.~14, No.~8, August~2015}%
{Kharma \MakeLowercase{\textit{et al.}}: Unveiling the Link: How Programming Languages Impact the Security and Quality of LLM-Generated Code}

\IEEEtitleabstractindextext{%
\begin{abstract}

Artificial Intelligence (AI) driven code generation tools are increasingly used throughout the software development lifecycle to accelerate coding tasks. However, the security of AI-generated code using large language models (LLMs) remains underexplored, and recent studies have revealed various risks and weaknesses. This paper presents a measurement study of LLM-generated code across four programming languages (Python, Java, C++, and C) and five widely used LLM families. We construct a manually curated dataset of 200 programming tasks, grouped into seven functional and security-relevant categories, each with language-neutral specifications. For every combination of task, language, and model, we generate code and evaluate it along three axes: syntactic validity and compilation success, semantic correctness using 4{,}000 per program unit test files, and software quality and security using SonarQube and CodeQL, complemented by manual review of key static analysis findings. Our results show clear language effects: Python and Java achieve higher compilation and semantic correctness rates and produce fewer security findings than C and C++, where we observe more memory safety issues, hard-coded secrets, and cryptographic misuses. We also find that many models fail to make use of modern security features available in recent compiler and toolkit updates (\ie in Java~17), and that outdated methods remain common, particularly in C++. These findings highlight the need to advance LLMs so that they better align with emerging secure coding practices and language-specific best practices. All code and data are available at \href{https://github.com/mohsystem/code-llm-evaluation-dataset.git}{GitHub}.

\end{abstract}

}

\maketitle

\IEEEdisplaynontitleabstractindextext

\IEEEpeerreviewmaketitle

\section{Introduction} \label{sec:introduction}

The rapid advancement of artificial intelligence (AI) technologies has driven the widespread adoption of large language models (LLMs) for generating source code across a broad range of programming tasks. In particular, LLMs have demonstrated remarkable performance in natural language understanding and generation. Building on these capabilities, recent LLMs have emerged as powerful tools for software developers, enabling the automated generation of functional code in multiple programming languages.

In 2018, Microsoft Visual Studio released the IntelliCode extension~\cite{IntelliC24}, which offers AI-powered development features that provide limited insights by analyzing code context using machine learning. In 2021, GitHub introduced Copilot, an AI-driven code assistant designed to improve coding quality by training on extensive real-world code repositories~\cite{IntroducCopilot24}, enabling it to provide coding recommendations across various programming languages and frameworks. Since GitHub introduced Copilot, AI-driven code generation adoption has grown in the software development lifecycle~\cite{AsareNA24}, where complex models are used to perform specific tasks like writing code for software engineers~\cite{ElgedawySDGGGJLLR24}. 

One major challenge for AI-driven code generation is ensuring code quality, where key metrics include validity, correctness, security, reliability, and maintainability. Current research stresses the importance of ensuring security in LLM-generated code~\cite{PerrySKB23,KhouryABC23}. Several other studies have shown that LLMs are capable of producing code that is both secure and vulnerable~\cite{Siddiq-2023,PerrySKB23,ElgedawySDGGGJLLR24}. The factors affecting the security of the code generated by these models are not identified. Multiple studies highlight the need for better user guidelines and awareness when interacting with AI tools~\cite{PerrySKB23,KhouryABC23,NairSM23} to enhance the quality properties of generated code. The security quality of the code generated for identical scenarios may differ depending on the chosen programming language, which is considered an explainable issue~\cite{RasXGD22,KhouryABC23}.

This work addresses this gap by investigating the factors that impact the security of LLMs-generated code, with a focus on the influence of language selection. Our contributions include the development of a dataset designed to evaluate AI-generated code across multiple domains, such as problem-solving, algorithms, and other key areas. Additionally, we evaluate the ability of various LLMs ({\em claude-3.5}, {\em gemini-1.5}, {\em codestral}, {\em GPT-4o}, {\em llama-3}) to generate secure and functional code in different languages (Python, Java, C++, and C). Moreover, We conduct a quality and security analysis of LLM-generated code using static security analysis tools (SAST) to identify common vulnerabilities and weaknesses, and quantify their distribution across models and languages, focusing on how the known CWE categories manifest in a unified evaluation setting. Lastly, this research conducts a comparative study of the security properties of AI-generated code across different programming languages, highlighting the main strengths and limitations of each AI tool in contexts such as semantic correctness and security.

\BfPara{Research Questions}
We address this problem through the following research questions.
\BfPara{RQ1} How does the choice of programming language affect the security of code generated by large language models (LLMs)?
\BfPara{RQ2} How do different LLMs compare in generating syntactically valid and semantically correct code across multiple programming languages?
\BfPara{RQ3} Which security vulnerabilities are most prevalent in LLM-generated code, and how do they vary across models and programming languages?

\BfPara{Contributions} We make the following contributions. (1) {\em New dataset for LLM-based coding.} We introduce a new manually vetted dataset of 200 programming tasks from seven categories that can be used by the research community for evaluating the performance of LLMs. 
\rev{(2) {\em Comparative evaluation of LLM-generated code.} We evaluate the quality and security of LLM-generated code across multiple programming languages and models using a unified dataset and a consistent evaluation setup. This enables direct cross-language and cross-model comparisons and exposes differences that are missed in single-language or single-model studies. We treat code quality and security as complementary dimensions, as low-quality code often correlates with security vulnerabilities and secure coding practices align with readability and maintainability. Joint evaluation establishes a baseline for analyzing their interactions and trade-offs in LLM-generated code.} \rev{(3) {\em In-depth security analysis.} We analyze LLM-generated code across multiple programming languages to identify language-specific characteristics, recurring issues, and distinct security vulnerability patterns.}


\BfPara{Organization} We provide a review of related work in \autoref{sec:relatedwork}, the research methodology is described in \autoref{sec:Methodology}, the proposed dataset is presented in \autoref{sec:Dataset_Description}, analysis results and discussion in \autoref{sec:AnalysisResults}, and the concluding remarks and future work in~\autoref{sec:Conclusion}.

\begin{table*}[t]
\caption{\normalfont A summary of the related work. Highlighted the evaluated programming languages and the LLMs. Languages: \ding{172} {\em C}, \ding{173} {\em C++}, \ding{174} {\em Java}, \ding{175} {\em Python}, \ding{176} {\em JavaScript}, \ding{177} {\em HTML}, \ding{178} {\em Verilog}. LLMs: \ding{179} {\em Copilot},  \ding{180} {\em Codex}, \ding{181} {\em Whisper}, \ding{182} {\em Gemini}, \ding{183} {\em Bard}, \ding{184} {\em GPTs}, \ding{185} {\em Llama-3}, \ding{186} {\em Claude-3.5}, \ding{187} {\em Codestral}, \ding{188} {\em StarCoder}, \ding{189} {\em CodeGen}.}\vspace{-3mm}
\scalebox{0.85}{
\begin{tabular}{lcccccccccccccccccccll}
 \hline
\multirow{2}{*}{\textbf{Reference}} &
\multirow{2}{*}{\textbf{Year}} &

\multicolumn{7}{c}{\textbf{Programming Languages}} & 
\multicolumn{1}{c}{\textbf{}} & 

\multicolumn{9}{c}{\textbf{Large Language Models}} &
&
&
\multirow{2}{*}{\textbf{Prompt Scenario}} 
\\
\cline{3-9} \cline{11-21}
  & & \ding{172} & \ding{173} & \ding{174}& \ding{175} &\ding{176} & \ding{177}& \ding{178}  &
& \ding{179}
& \ding{180} 
& \ding{181} 
& \ding{182} 
& \ding{183} 
& \ding{184} 
& \ding{185} 
& \ding{186}
& \ding{187}
& \ding{188} 
& \ding{189}\\ 
 \hline

 Asare \etal\cite{AsareNA24} & 2024& \ding{52} & \ding{22} & \ding{22}&\ding{22}&\ding{22}&\ding{22}&\ding{22}
 & & \ding{52}&\ding{22}&\ding{22}&\ding{22}&\ding{22}&\ding{22}&\ding{22}&\ding{22}&\ding{22}&\ding{22}&\ding{22}&Two problems~\cite{copilotUSS24}\\

Perry \etal\cite{PerrySKB23}& 2023& \ding{52} & \ding{22} & \ding{22}&\ding{52}&\ding{52}&\ding{22}&\ding{22}
 & & \ding{22}&\ding{52}&\ding{22}&\ding{22}&\ding{22}&\ding{22}&\ding{22}&\ding{22}&\ding{22}&\ding{22}&\ding{22}&Six tasks~\cite{PerrySKB23}\\

  Sandoval \etal\cite{SandovalPNKGD23} & 2023& \ding{52} & \ding{22} & \ding{22}&\ding{22}&\ding{22}&\ding{22}&\ding{22}
 & & \ding{22}&\ding{52}&\ding{22}&\ding{22}&\ding{22}&\ding{22}&\ding{22}&\ding{22}&\ding{22}&\ding{22}&\ding{22}&Shopping list function\\

 Asare \etal\cite{AsareNA23}  & 2023& \ding{52} & \ding{52} & \ding{22}&\ding{22}&\ding{22}&\ding{22}&\ding{22}
 & & \ding{52}&\ding{22}&\ding{22}&\ding{22}&\ding{22}&\ding{22}&\ding{22}&\ding{22}&\ding{22}&\ding{22}&\ding{22}&Big-Vul dataset~\cite{FanLWN20}\\

   Yetistiren \etal\cite{YetistirenOAT23} & 2023& 
\ding{22}&\ding{22}&\ding{22}&\ding{52}&\ding{22}&\ding{22}&\ding{22}
 & & \ding{52}&\ding{22}&\ding{52}&\ding{22}&\ding{22}&\ding{52}&\ding{22}&\ding{22}&\ding{22}&\ding{22}&\ding{22}& HumanEval dataset~\cite{ChenTJYPKEBJBRPKPKSMCGRPPKBWTSCPCBHGNPTTBBJSHC21}\\

   Khoury \etal\cite{KhouryABC23} & 2023& \ding{52} & \ding{52} & \ding{52}&\ding{52}&\ding{22}&\ding{52}&\ding{22}
 & & \ding{52}&\ding{22}&\ding{22}&\ding{22}&\ding{22}&\ding{22}&\ding{22}&\ding{22}&\ding{22}&\ding{22}&\ding{22}&21 tasks~\cite{RaphaelK24}\\

   Nair \etal\cite{NairSM23} & 2023& \ding{22} & \ding{22} & \ding{22}&\ding{22}&\ding{22}&\ding{22}&\ding{52}
 & & \ding{52}&\ding{22}&\ding{22}&\ding{22}&\ding{22}&\ding{22}&\ding{22}&\ding{22}&\ding{22}&\ding{22}&\ding{22}&Scenarios from the selected CWEs\\

  Elgedawy \etal\cite{ElgedawySDGGGJLLR24} & 2023& \ding{22} & \ding{22} & \ding{22}&\ding{52}&\ding{22}&\ding{22}&\ding{22}
 & & \ding{22}&\ding{22}&\ding{22}&\ding{52}&\ding{52}&\ding{52}&\ding{22}&\ding{22}&\ding{22}&\ding{22}&\ding{22}&Nine tasks\\

  Wu \etal\cite{WuZBBZWX23} & 2023& \ding{22} & \ding{22} & \ding{52}&\ding{52}&\ding{22}&\ding{22}&\ding{22}
 & & \ding{22}&\ding{22}&\ding{22}&\ding{22}&\ding{22}&\ding{52}&\ding{22}&\ding{22}&\ding{22}&\ding{22}&\ding{22}&SARD and Juliet datasets~\cite{Black-2018}\\

  Siddiq \etal\cite{Siddiq-2023} & 2023& \ding{22} & \ding{22} & \ding{22}&\ding{52}&\ding{22}&\ding{22}&\ding{22}
 & & \ding{22}&\ding{22}&\ding{22}&\ding{22}&\ding{22}&\ding{52}&\ding{22}&\ding{22}&\ding{22}&\ding{52}&\ding{52}&LLMSecEval dataset~\cite{LLMSecEvalDS24}\\
 
  Schuster \etal\cite{SchusterSTS21} & 2021& \ding{22} & \ding{22} & \ding{22}&\ding{52}&\ding{22}&\ding{22}&\ding{22}
 & & \ding{22}&\ding{22}&\ding{22}&\ding{22}&\ding{22}&\ding{52}&\ding{22}&\ding{22}&\ding{22}&\ding{22}&\ding{22}&-\\
  Ullah \etal\cite{ullah2024llms} & 2024& \ding{52} & \ding{52} & \ding{52}&\ding{52}&\ding{22}&\ding{22}&\ding{22}
 & & \ding{22}&\ding{22}&\ding{22}&\ding{22}&\ding{22}&\ding{22}&\ding{52}&\ding{22}&\ding{52}&\ding{52}&\ding{52}& 200 tasks\\
 Siddiq \etal\cite{SiddiqS22} & 2022& \ding{22} & \ding{22} & \ding{22}&\ding{52}&\ding{22}&\ding{22}&\ding{22}
 & & \ding{52}&\ding{22}&\ding{22}&\ding{22}&\ding{22}&\ding{22}&\ding{22}&\ding{22}&\ding{22}&\ding{22}&\ding{22}&SecurityEval dataset~\cite{LLMSecEvalDS24}\\
 Khare \etal\cite{khare2025understanding} & 2024& \ding{52} & \ding{52} & \ding{52}&\ding{22}&\ding{22}&\ding{22}&\ding{22}
 & & 
 \ding{52}&\ding{22}&\ding{52}&\ding{52}&\ding{22}&\ding{52} &\ding{22}&\ding{22}&\ding{22}&\ding{22}& \ding{22}& 25 CWEs, dataflow based prompt\\
  Asare \etal\cite{asare2024llms} & 2024& \ding{52} & \ding{52} & \ding{52}&\ding{52}&\ding{22}&\ding{22}&\ding{22}
 & & \ding{22}&\ding{22}&\ding{22}&\ding{22}&\ding{22}&\ding{52}&\ding{22}&\ding{52}&\ding{22}&\ding{52}&\ding{22}& 200 Tasks\\
 Tony \etal\cite{tony2023llmseceval} & 2023& \ding{52} & \ding{22} & \ding{22}&\ding{52}&\ding{22}&\ding{22}&\ding{22}
 & & \ding{22}&\ding{22}&\ding{22}&\ding{52}&\ding{22}&\ding{22}&\ding{22}&\ding{52}&\ding{22}&\ding{52}&\ding{22}& LLMSecEval dataset \cite{LLMSecEvalDS24}\\
 Lenarduzzi \etal\cite{lenarduzzi2023critical} & 2021& \ding{22} & \ding{22} & \ding{52}&\ding{52}&\ding{22}&\ding{22}&\ding{22}
 & & \ding{22}&\ding{22}&\ding{22}&\ding{52}&\ding{22}&\ding{22}&\ding{22}&\ding{22}&\ding{22}&\ding{22}&\ding{22}& LLMSecEval dataset \cite{LLMSecEvalDS24}\\
\xl{1}
  This work & 2024& \ding{52} & \ding{52} & \ding{52}&\ding{52}&\ding{22}&\ding{22}&\ding{22}
 & & \ding{22}&\ding{22}&\ding{22}&\ding{52}&\ding{22}&\ding{52}&\ding{52}&\ding{52}&\ding{52}&\ding{22}&\ding{22}& 200 tasks\\

\xl{2}
\end{tabular}}
\label{Table:related_work}\vspace{-4mm}
\end{table*}

\section{Related Work} \label{sec:relatedwork}

Several works explored LLMs's use in a variety of software development activities~\cite{HuangCCCPTHXZ24,PerrySKB23,SandovalPNKGD23,AsareNA24,YetistirenOAT23,AsareNA23,KhouryABC23,NairSM23,ElgedawySDGGGJLLR24,Siddiq-2023,SchusterSTS21,WuZBBZWX23,ullah2024llms, PearceATDK22, SiddiqS22, tony2023llmseceval, lenarduzzi2023critical}. This section examines key studies, emphasizing their methods, applications, and the analyzed features. The comparative overview helps place our work in the broader research context.

Huang \etal~\cite{HuangCCCPTHXZ24} surveyed pre-trained models and LLMs for software engineering tasks, including requirements, code, test case, patch generation, optimization, summarization, and code translation, covering models such as {\em BERT}, general transformers, and {\em ChatGPT}. Our work builds on this line of research by focusing specifically on code generation and evaluating the effectiveness of LLMs in this setting. Perry \etal~\cite{PerrySKB23} analyzed the security of LLM-generated code and showed that it often contains more security flaws than human-written code, with LLMs exhibiting overconfidence in code safety. Sandoval \etal~\cite{SandovalPNKGD23} reported a 10\% increase in vulnerabilities in LLM-generated C code. Asare \etal~\cite{AsareNA24} found that {\em GitHub Copilot} can improve security for complex tasks but provides limited benefit for simpler ones. In contrast to prior work, we evaluate four LLMs across multiple programming languages, including C++ and Java, to enable broader cross-language and cross-model analysis.

Yeti\c{s}tiren \etal~\cite{YetistirenOAT23} evaluated the quality of code generated by three code assistants using the HumanEval benchmark~\cite{ChenTJYPKEBJBRPKPKSMCGRPPKBWTSCPCBHGNPTTBBJSHC21}. They reported correctness rates of 65.2\%, 46.3\%, and 31.1\% for {\em ChatGPT}, {\em Copilot}, and {\em CodeWhisperer}, respectively, and analyzed the impact of function names, inputs, and problem descriptions. Security, maintainability, and reliability were assessed using SonarQube~\cite{SonarQube24}, with all tools producing code deemed secure. Our work differs by evaluating four programming languages and using a broader and more diverse set of dataset scenarios.

Asare~\etal\cite{AsareNA23} compared {\em GitHub Copilot's} code generation to human-written code for security vulnerabilities, using a dataset by Fan~\etal~\cite{FanLWN20}. They found that {\em Copilot} recreated the same vulnerabilities in 33.3\% of cases and remedied 25.5\% of them. {\em Copilot} showed inconsistencies, especially with older vulnerabilities, but generally produced fewer security flaws than humans. Khoury~\etal\cite{KhouryABC23} tested {\em ChatGPT's} ability to generate secure code in various languages, showing that it often failed to meet security standards but improved with follow-up prompts. {\em ChatGPT} corrected 12 of the 21 programs when prompted. 

Nair \etal\cite{NairSM23} explored {\em ChatGPT's} effectiveness in generating hardware code using Common Vulnerability Enumerations (CWE-1194), showing the possibility of guiding AI to avoid common flaws. Elgedawy~\etal\cite{ElgedawySDGGGJLLR24} analyzed {\em GPT-3.5}, {\em GPT-4}, {\em Bard}, and {\em Gemini}, showing that using security personas reduced vulnerabilities, especially in {\em GPT-3.5}, {\em GPT-4}, and {\em Bard}. Siddiq~\etal\cite{Siddiq-2023} introduced the SALLMS framework to evaluate LLMs systematically for security. Schuster \etal\cite{SchusterSTS21} and Wu \etal\cite{WuZBBZWX23} highlighted challenges, such as data poisoning and the reduced effectiveness of LLMs in handling complex security issues, highlighting the need for stronger defenses and better vulnerability detection.

Ullah \etal~\cite{ullah2024llms} evaluated modern LLMs for vulnerability detection and identified limitations in consistency, reasoning accuracy, and robustness, particularly under minor code changes and real-world conditions. Pearce \etal~\cite{PearceATDK22} analyzed the security implications of code generated by {\em GitHub Copilot} and showed that the tool frequently produces vulnerable patterns, especially when prompts are underspecified or ambiguous. Siddiq and Santos~\cite{SiddiqS22} introduced SecurityEval, a dataset of 130 Python programs mapped to 75 common weakness enumerations (CWEs), to evaluate the security of machine-generated code. Using SecurityEval, they assessed tools such as {\em GitHub Copilot} and {\em InCoder} and found that these models often generate code with security vulnerabilities, emphasizing the need to evaluate generated code beyond functional correctness to include security.

Asare \etal~\cite{AsareNA23} evaluated the ability of LLMs to detect and reason about security vulnerabilities in source code across multiple languages and tasks. The study showed that most LLMs fail to identify deeper security flaws, even when the code is syntactically correct, highlighting the need for more reliable, security-focused benchmarks and evaluation methods. Tony \etal~\cite{tony2023llmseceval} introduced LLMSecEval, a dataset of natural language prompts designed to assess LLM performance on security-related questions across common software security topics, enabling consistent evaluation of security understanding. Together, these studies stress the importance of evaluating LLMs beyond code generation, with explicit focus on security reasoning.

Lenarduzzi \etal~\cite{lenarduzzi2023critical} compared six code analysis tools and evaluated their issue detection effectiveness, result consistency, and precision and reported substantial differences in detection coverage and overlap among tools, showing that reliance on a single tool can miss vulnerabilities or introduce false positives. Khare \etal~\cite{khare2025understanding} performed a large-scale evaluation of LLM-based vulnerability detection, testing 16 models on 5{,}000 code samples from five datasets of C, C++, and Java. The authors proposed dataflow-inspired prompting strategies and showed that LLMs can outperform static and deep learning-based tools for specific vulnerability classes while providing interpretable explanations.

\autoref{Table:related_work} summarizes prior work by language, LLM, and prompt for LLM-generated code. \autoref{Table:related_work_quality} categorizes related work by the quality attributes examined and the security analysis methods applied to evaluate generated code, and reports the overall impact of AI code generation on code quality based on each study’s experimental results.

\BfPara{Our Work}
Most prior studies analyze LLM code generation behavior but provide limited insight into the factors affecting code quality and security beyond user demographics and experience levels. This work examines how programming language characteristics influence the quality of LLM-generated code. We evaluate validity, correctness, security, maintainability, consistency, intentionality, adaptability, and responsibility across four languages and five LLMs. This study is a quantitative measurement study: it introduces no new vulnerability classes or analysis techniques, but systematically compares the behavior of existing LLMs across languages and tasks under a unified setup.

\begin{table}[t]
\setlength\extrarowheight{-1pt}
\caption{\normalfont Summary of related work. Overall Impact (OI): Negative (N), Positive (P), and Negative due to the use of an inappropriate dataset for security (N*). Attributes: \underline{V}alidity, \underline{C}orrectness, \underline{Se}curity, \underline{Re}liability, and \underline{Ma}intainability. Methods: \underline{M}anual, \underline{S}tatic Scan, and \underline{R}untime Scan.}\vspace{-3mm}
\scalebox{0.80}{
\begin{tabular}{l|cccccc|ccc|l}
\hline
\multirow{2}{*}{\textbf{Reference}} &
\multicolumn{6}{c|}{\textbf{Quality Attributes}} &
\multicolumn{3}{c|}{\textbf{Analysis Method}} &
\multirow{2}{*}{\textbf{OI}} \\
\cline{2-7} \cline{8-10}
& 
V& 
C & 
Se& 
Re& 
Ma& 
Pe&
M & 
S & 
R\\ 
 \hline

 Asare \etal\cite{AsareNA24}&\ding{22}&\ding{22} & \ding{52}&\ding{22}&\ding{22}&\ding{22}&\ding{52} & \ding{22}&\ding{22}&N\\

Perry \etal\cite{PerrySKB23}&\ding{22}&\ding{22} & \ding{52}&\ding{22}&\ding{22}&\ding{22}&\ding{52} & \ding{22}&\ding{22}&N\\

  Sandoval \etal\cite{SandovalPNKGD23}&\ding{22}&\ding{22} & \ding{52}&\ding{22}&\ding{22}&\ding{22}&\ding{52} & \ding{52}&\ding{52}&N\\

 Asare \etal\cite{AsareNA23}&\ding{22}&\ding{22} & \ding{52}&\ding{22}&\ding{22}&\ding{22}&\ding{52} & \ding{52}&\ding{22}&P\\

   Yetistiren \etal\cite{YetistirenOAT23}&\ding{52}&\ding{52} & \ding{52}&\ding{52}&\ding{52}&\ding{22}&\ding{52} & \ding{52}&\ding{22}&N*\\

   Khoury \etal\cite{KhouryABC23}&\ding{22}&\ding{22} & \ding{52}&\ding{22}&\ding{22}&\ding{22}&\ding{52} & \ding{22}&\ding{22}&N\\

   Nair \etal\cite{NairSM23}&\ding{22}&\ding{22} & \ding{52}&\ding{22}&\ding{22}&\ding{22}&\ding{52} & \ding{22}&\ding{22}&P\\

  Elgedawy \etal\cite{ElgedawySDGGGJLLR24}&\ding{52}&\ding{52} & \ding{52}&\ding{52}&\ding{22}&\ding{52}&\ding{52} & \ding{52}&\ding{22}&N\\

  Wu \etal\cite{WuZBBZWX23}&\ding{22}&\ding{22} & \ding{52}&\ding{22}&\ding{22}&\ding{22}&\ding{52} & \ding{22}&\ding{22}&N\\
 Siddiq \etal\cite{Siddiq-2023} & \ding{22} & \ding{22} & \ding{52} & \ding{22} & \ding{22} & \ding{22} & \ding{22} & \ding{52} & \ding{22} & N\\

 Khare \etal\cite{khare2025understanding} & \ding{52} & \ding{52} & \ding{52} & \ding{22} & \ding{52} & \ding{52} & \ding{22} & \ding{52} & \ding{22} & P\\

 Asare \etal\cite{AsareNA24} & \ding{22} & \ding{22} & \ding{52} & \ding{22} & \ding{22} & \ding{22} & \ding{52} & \ding{22} & \ding{22} & N\\

 Tony \etal\cite{tony2023llmseceval} & \ding{22} & \ding{22} & \ding{52} & \ding{22} & \ding{22} & \ding{22} & \ding{22} & \ding{52} & \ding{22} & N\\

 Lenarduzzi \etal\cite{lenarduzzi2023critical} & \ding{22} & \ding{22} & \ding{52} & \ding{22} & \ding{22} & \ding{22} & \ding{22} & \ding{52} & \ding{22} & N\\
 Ullah \etal\cite{ullah2024llms} & \ding{22} & \ding{22} & \ding{52} & \ding{22} & \ding{22} & \ding{22} & \ding{22} & \ding{52} & \ding{22} & N\\

  Siddiq \etal\cite{Siddiq-2023}&\ding{22}&\ding{22} & \ding{52}&\ding{22}&\ding{22}&\ding{22}&\ding{22} & \ding{52}&\ding{22}&N\\

  Schuster \etal\cite{SchusterSTS21}&\ding{22}&\ding{22} & \ding{52}&\ding{22}&\ding{22}&\ding{22}&\ding{52} & \ding{22}&\ding{22}&N\\
 \xl{1}
  This work &\ding{52}&\ding{52} & \ding{52}&\ding{52}&\ding{52}&\ding{22}&\ding{52} & \ding{52}&\ding{22}&N\\

\xl{2}
\end{tabular}}
\label{Table:related_work_quality}\vspace{-4mm}
\end{table}

\section{Methodology}\label{sec:Methodology}

We selected LLMs using a two-step process. First, we defined eligibility criteria based on prior literature and industry benchmarks: \ding{172} widespread adoption in software engineering practice, \ding{173} demonstrated performance in program synthesis, reasoning, and security-related tasks, \ding{174} official support for code generation in Python, Java, C++, and C, \ding{175} stable and accessible APIs during data collection, and \ding{176} architectural diversity to avoid evaluating closely related model variants. Second, we applied these criteria to available models and list the selected LLMs in~\autoref{tab:selected_models}. The selected models are widely used, frequently evaluated in prior studies, support all target languages through stable APIs, and represent distinct training and design choices, enabling systematic comparative analysis.

As is known, the internal architecture and training methodologies of these LLMs,~\ie can introduce significant variation in the code that is produced. This study aims to provide information on how these variations influence the security posture of generated code, focusing on the importance of selecting the right LLM based on specific development needs.~\autoref{tab:model_configurations} highlights the context window and other configurations used when generating code using each model. Max\_tokens refers to the maximum number of tokens to generate in completion. Top\_p changes how the model selects tokens for output. Tokens are selected from the most probable to least until the sum of their probabilities equals the top-p value. The model temperature is used to control the randomness in generating the output. The context window is the maximum token count a model can handle in one forward pass, covering both the input (prompt) and the output. It essentially determines the amount of text the model can process at a time. 

\begin{table}[t]
\centering
\caption{\normalfont LLMs used for evaluation and their short names.}\label{tab:selected_models}\vspace{-3mm}
\scalebox{0.85}{
\begin{tabular}{llll}
\xl{2}
{Provider}& { Ref} & {Model~\rev{ Version}} & { Short}                                           \\
\xl{1}
OpenAI &\cite{GPT4Open24}        & GPT-4o\rev{-202407} &{\em GPT-4o}           \\
Perplexity &\cite{Perplexity}         & llama-3-sonar-large-32k-chat  & {\em llama-3}              \\
\rev{Anthropic} &\cite{Claude}          & claude-3-5-sonnet-20240620   & {\em claude-3.5}         \\
Mistral~\rev{AI}&\cite{Mistral24}  & codestral-\rev{2405}  &  {\em codestral}           \\
Google&\cite{Gemini24}       & gemini-1.5-pro-001 & {\em gemini-1.5}            \\
\xl{2}
\end{tabular}
}
\end{table}

\begin{table}[t]
\centering
\caption{\normalfont LLMs, temperature (Temp), maximum tokens (MaxT), context window (CW), and {Top\_P (TopP)}.}\label{tab:model_configurations}\vspace{-3mm}\scalebox{0.95}{
\begin{tabular}{lrrrc}
\xl{2}
{ Model} & {Temp} & {MaxT} & {TopP}  & {CW}\\
\xl{1}
{\em GPT-4o} & 0.9 & 4,096 & 0.9 & 128k \\
{\em llama-3} & 0.9 & 4,096 & 0.9 & 32k \\
{\em claude-3.5} & 0.9 & 4,096 & 0.9 & 200k \\
{\em codestral} & 0.9 & 4,096 & 0.9 & 32k\\
{\em gemini-1.5} & 0.9 & 4,096 & 0.9 & 128k \\
\xl{2}
\end{tabular}
}\vspace{-4mm}
\end{table}

\subsection{Programming Languages Selection} One of the key motivations and contributions of this study is the comparative exploration of the performance and security of LLMs under the same evaluation settings for different programming languages. We determine a range of programming languages to be evaluated in our evaluation, covering both statically and dynamically typed languages. As such, we choose the following programming languages as a preliminary set that meets those metrics: \ding{172} 
 C; \ding{173} C++; \ding{174} Java; \ding{175} and Python. Although our choice of programming language is limited by the capabilities of LLMs and the programming languages they support, we believe that these programming languages are representative, so they are among the top five most used programming languages~\cite{MostPopuPL24}. 

Each programming language exhibits attributes that affect security outcomes in code generation (e.g., Python’s dynamic typing versus C++’s static typing leads to different classes of bugs and vulnerabilities). Java and Python provide automatic memory management, reducing memory-related errors relative to C and C++, which rely on manual memory management. These languages therefore enable analysis of how LLMs handle language-specific security properties. The selection also improves external validity, as these languages rank among the top five most used in 2024 and reflect common real-world development settings.

\subsection{Dataset}
To comprehensively evaluate LLM performance, we curated 200 prompt descriptions targeting diverse aspects of code generation. Task selection ensured broad coverage of core programming paradigms and secure coding practices. Details are provided in \autoref{sec:Dataset_Description}.

\subsection{Environment Setup}
To ensure consistency and reproducibility, we used a standardized environment for code generation, compilation, execution, and validation across tasks and programming languages, including unit test support. All experiments ran on a Lenovo ThinkPad E570 with a 7th-generation Intel\textsuperscript{\textregistered} Core\textsuperscript{\texttrademark} i7 processor, 16\,GB DDR4 RAM, and a 256\,GB SSD. This configuration provides sufficient performance and portability for LLM integration, code generation, and multi-language compilation. We used Debian~12 for its stability, package management, and low resource overhead, ensuring a consistent platform for cross-language testing. The following software packages and versions were used. \ding{182} \BfPara{Java} We compiled and executed Java code using OpenJDK~17.0.8 (LTS), which supports modern Java language features and ensures compatibility with constructs generated by LLMs.
\ding{183} \BfPara{Python} We evaluated Python code using Python~3.11.9 to ensure compatibility with current features and libraries.
\ding{184} \BfPara{C and C++} We compiled C and C++ code using CMake~3.28.6. For C++, we set \texttt{CMAKE\_CXX\_STANDARD} to~17, a widely adopted standard since its 2017 release~\cite{CppProgramSurvey}, ensuring support for modern C++ features produced by LLMs.

This setup offered a stable basis for assessing the code produced by the five chosen LLMs. By preserving identical hardware and software settings, we ensured that any differences in code compilation, run-time, or accuracy between programming languages or models are due to the models themselves and not to environmental factors.

\subsection{LLM Integration and Code Generation}\label{sec:llm_integration_generation}
Based on the LLMs listed in~\autoref{tab:selected_models}, we generated code for each task in four programming languages. A custom Python script sequentially dispatched task prompts to each model and collected the generated code. All inputs used the original task descriptions without prompt engineering. After generation, a second Python script parsed and organized the outputs into a structured file system, assigning language-specific file extensions (e.g., \texttt{.py} for Python and \texttt{.java} for Java). Each file name encoded the task identifier, LLM, and programming language to support traceability and comparison. The first author verified the total number of generated programs to ensure complete collection.

As a result, 4,000 code files were generated, comprising 200 files per language for each LLM. This dataset formed the foundation for the subsequent evaluation procedures, ensuring that each LLM was evaluated using an identical set of tasks and programming languages.

\subsection{Quality Evaluation}
To evaluate code quality, we analyze syntax validity, functional correctness, code length, reliability, maintainability, and security~\cite{YetistirenOAT23, Siddiq-2023} using two evaluation approaches. \ding{172} \emph{Manual evaluation} by human experts where two software engineers with four and ten years of professional experience participated in the evaluation, with the first and second authors coordinating and reviewing the process. Drafting, validating, and executing test cases to assess semantic correctness required over 800 person-hours across five months. \ding{173} \emph{Automated static analysis} using SonarQube~\cite{SonarQube24} and CodeQL~\cite{CodeQL}, which support all target languages. Automated tools provide scalable coverage of quality and security metrics, but have known limitations when used in isolation. Human evaluation remains the gold standard for assessing LLM outputs. We therefore complement automated analysis with human evaluation on a representative subset of generated programs to support generalization beyond a limited number of evaluators. Prior work identifies these methods as effective for code quality and security assessment, supporting our choice of metrics and tools.

\subsubsection {Semantic Soundness}
\label{sec:Semantic_Soundness}
\BfPara{Compilation-Time Errors} Before evaluating functionality and quality, we first ensured the code met syntactic standards and compiled without errors. This step was crucial for valid and reliable assessments.

For each programming language, automated scripts were created to perform syntax checks and were reviewed by one of the software engineers and the second author. For compiled languages (C, C++, and Java), each source file was compiled and the outcome recorded. For interpreted languages like Python, syntax validation was conducted using Python's built-in checker. Results were documented in a matrix of 200 rows (tasks) by 20 columns (5 LLMs $\times$ 4 languages), where each cell contained a binary value indicating syntactic validity or successful compilation. This process enabled early identification of issues such as syntax errors or missing imports, allowing for the correction or exclusion of problematic samples prior to semantic analysis.

\BfPara{\rev{Semantic Correctness Using Unit Tests}}
We evaluated semantic correctness after resolving all syntactic errors by checking whether each generated program implemented the intended logic specified in the task prompt.

The initial goal was to create a single reusable unit test per task and language. Natural language prompts without fixed function signatures led to substantial variation in generated code, including differences in function names, parameters, return values, and I/O handling. This variability reflects realistic LLM use but prevented uniform test reuse.

We addressed this variability using a four-phase unit testing procedure. First, for each of the 200 tasks, we defined a language-agnostic specification describing intended behavior, input domains, and test scenarios, including normal, boundary, edge, and invalid cases. Reference implementations produced expected outputs. Second, after code generation, we analyzed each program’s interface, including function signatures, parameter types and order, output mechanisms, and data structures (e.g., Python lists versus \texttt{std::vector} in C++), which ruled out a unified test suite.

Third, we implemented per-sample unit tests tailored to each program's structure while preserving the shared logical specification. All tests used the same inputs and expected outputs across languages and models, adapting only syntax and data types. Assertions targeted return values or captured standard output based on program behavior. Fourth, two professionals manually reviewed each test for correctness, and the first and second authors re-checked a representative subset across languages and task categories. This process enabled consistent and fair evaluation of semantic correctness despite structural differences in generated code.



In total, we created 4{,}000 unit test files, corresponding to 200 tests per programming language per LLM. Unit test outcomes were recorded in a matrix with the same dimensions used for compilation-time error analysis. \rev{Each unit test received a score from 0\% (no tests passed) to 100\% (all tests passed), based on the proportion of successful test cases. Two developers executed the testing, and the first and second authors conducted a single review cycle of the unit test files. This process quantified semantic correctness and reflected how accurately LLMs interpreted task specifications. The evaluation posed challenges, including inconsistent function signatures across LLM outputs, which prevented standardized testing. We addressed this by creating per-sample unit tests, increasing implementation effort but ensuring correctness. Additional variability in code quality and completeness required limited manual fixes, such as adding missing imports, to enable compilation and testing.}

Beyond semantics, we analyzed code coverage on a representative subset of tasks using JaCoCo~\cite{jacocoja75} for Java and comparable tools for other languages. Each task included up to 10 unit tests covering standard cases and edge conditions. Coverage analysis complemented correctness checks by assessing test thoroughness. Average line coverage across sampled tasks exceeded 75\%, indicating that most logic paths in the generated code were exercised. This coverage level supports the robustness of test design and evaluation.

\subsubsection{Complexity}
\rev{Understanding code complexity is essential for assessing program maintainability and readability. Higher complexity increases the likelihood of bugs, complicates testing, and lowers developer productivity. In this work, we analyze established complexity metrics to capture various structural and cognitive aspects of difficulty in LLM-generated code.} 
\begin{enumerate*}[leftmargin=*]
    \item[\ding{172}] {\em Lines of Code (LoC)} measures the program's lines of code, excluding whitespace. LoC is a predictive metric used to evaluate effort and maintainability.

    \item[\ding{173}] {\em Cyclomatic Complexity (CyC)}  calculates code complexity using a control flow graph (CFG). With \( E \) as edges, \( N \) as nodes, and \( Q \) as connected components, CyC is computed as \( M = E + 2Q - N \)~\cite{WatsonWM96}.

    \item[\ding{174}] {\em Cyclomatic Complexity Density (CCD)} measures how cyclomatic complexity spreads across the codebase. With \( cl \) as the total code lines, CCD is calculated as \( Md = {(E + 2Q - N)}/{cl} \)~\cite{FentonN99}.

    \item[\ding{175}] {\em Cognitive Complexity (CoC)} measures the difficulty of understanding code~\cite{BaronWW20}. CoC considers structure like control flow and nesting, using \( C = C_\text{base} + \sum_{i=1}^{n} {nc} \), where \( nc \) represents increases due to nesting and conditionals.

\end{enumerate*}

\subsubsection{Quality and Security}
\label{meth_quality_security}

We evaluate software quality and security using SonarQube~\cite{SonarQube24} and CodeQL~\cite{CodeQL}, two widely adopted static analyzers supporting all target languages. SonarQube is used to assess non-vulnerability quality dimensions, while CodeQL provides query-based validation of vulnerabilities. Consistent with prior work, we manually reviewed all reported vulnerabilities and validated a stratified 20\% sample across languages and LLMs to mitigate false positives and contextual limitations. Our evaluation covers four indicators:  reliability,  security (vulnerabilities),  maintainability, and security hotspots. In total, the analysis spans 97{,}412 lines of code across four languages by five LLMs. \ding{172} {\em Reliability} assesses correct operation under predefined conditions using reported bug findings.
\ding{173} {\em Security} captures exploitable weaknesses, quantified by the number and severity of detected vulnerabilities.
\ding{174} {\em Maintainability} measures ease of understanding and modification based on code smells and complexity indicators.
\ding{175} {\em Security Hotspots} identify security-sensitive code regions requiring careful handling, such as authentication, authorization, and input validation.

\subsubsection{Clean Code}
\rev{We evaluate clean code in LLM-generated outputs using four dimensions derived from SonarQube’s static analysis: consistency, intentionality, adaptability, and responsibility.}

\begin{enumerate}[leftmargin=*]
\item[\ding{172}] {\em Consistency} assesses formatting, naming conventions, and structural uniformity across languages and prompts, including spacing, indentation, and identifier casing. Consistent code improves readability and reduces review and maintenance overhead. \rev{SonarQube captures consistency issues via code smells such as improper naming, unused imports or dead code, and inconsistent indentation or braces, which are aggregated under maintainability due to their impact on code clarity.}

\item[\ding{173}] {\em Intentionality} evaluates whether the code clearly and efficiently implements the intended functionality without unnecessary complexity. \rev{SonarQube captures this attribute through cognitive complexity metrics and code smells that penalize deep nesting, large switch blocks, unclear logic, and bloated methods, directly contributing to maintainability.}

\item[\ding{174}] {\em Adaptability} measures ease of modification through modular design and separation of concerns, enabling localized changes with minimal risk. \rev{SonarQube assesses adaptability using function length, parameter count, and lack of modular decomposition (e.g., monolithic blocks), which affect maintainability and may trigger security hotspots when sensitive logic lacks encapsulation.}

\item[\ding{175}] {\em Responsibility} reflects compliance with professional and ethical standards, including secure data handling and avoidance of high-risk practices. \rev{We evaluate responsibility through detected vulnerabilities (e.g., hard-coded credentials), security hotspots (e.g., unchecked inputs or return values), and use of deprecated or insecure APIs, spanning both security and maintainability dimensions.}
\end{enumerate}

\BfPara{Analysis}
We analyze the results across key quality attributes, including syntactic validity, semantic correctness, security, reliability, maintainability, and clean code properties. Detailed results appear in~\autoref{sec:AnalysisResults}.

To improve methodological transparency, \autoref{fig:qc-pipeline} summarizes the complete Quality Control (QC) pipeline applied to each task, covering dataset design, code generation, compilation, semantic testing, and static analysis.

\begin{figure}[t]
\centering
\begin{tikzpicture}[
    node distance=3.9mm,
    >={Latex},
    font=\scriptsize,
    proc/.style={rectangle, draw=black!70, fill=black!8, rounded corners=3pt, 
                 align=left, text width=8.5cm, inner sep=4pt, line width=0.7pt, dashed}
]

\node[proc] (step1) {\BfPara{1) Dataset} 
Dataset construction, tagging, difficulty labels, language-neutral behavioral specification and test scenarios.\\
\BfPara{QC} tasks, tags, and specs were peer-reviewed by two (professional) software engineers working alongside
the first/second authors of this study. (Sec.~\ref{sec:dataset_Construction},  Sec.~\ref{sec:Semantic_Soundness}).};

\node[proc, below=of step1] (step2) {\BfPara{2) LLM code generation} 
For each task and language, five LLMs generate code under scripted prompts, outputs stored in structured repositories.\\
\BfPara{QC} scripted calls, fixed prompts, and traceable file naming ensure reproducibility, the quantity of generated programs was verified again by the first author to ensure that no LLM-produced output was missed (Sec.~\ref{sec:llm_integration_generation}).};

\node[proc, below=of step2] (step3) {\BfPara{3) Compilation} 
Language-specific scripts compile/parse each file, compilation matrix recorded. Minor, consistent fixes (imports/includes) applied and recorded when aligned with the prompt.\\
\BfPara{QC} compilation failures and fixes inspected, logged, and reviewed by one of the software engineers and the second author. (Sec.~\ref{sec:Semantic_Soundness}).};

\node[proc, below=of step3] (step4) {\BfPara{4) Program unit tests} 
Using shared unit test specification, one unit test file per program, adapted to its interface (signatures, parameter order, data structures, return vs.\ print) while keeping identical logical cases.\\
\BfPara{QC} each test file was manually
reviewed by two professionals to verify that it exercised the
intended functional behavior, and a representative subset
across languages and task categories was re-checked by
the first and second authors (Sec.~\ref{sec:Semantic_Soundness}, App.~\ref{APPENDIX_unit_test_example}).};

\node[proc, below=of step4] (step5) {\BfPara{5) Execution and semantic scoring}
Language-specific environment is used to execute all tests, per-program 0 - 100\% semantic score computed from passed cases, coverage sampled for selected tasks.\\
\BfPara{QC} the process was carried out by two developers, with one review cycle for the written unit test files conducted by the first and second authors of this work. Summary matrices are inspected for outliers, unexpected failures trigger manual re-check of tests and specification (Sec.~\ref{sec:Semantic_Soundness}).};

\node[proc, below=of step5] (step6) {\BfPara{6) Static analysis}
SonarQube evaluates reliability, security, maintainability, and hotspots, in addition to CodeQL using official query packs.\\
\BfPara{QC} unified configuration across languages, reports exported for manual inspection (Sec.~\ref{meth_quality_security}).};

\node[proc, below=of step6] (step7) {\BfPara{7) Cross-tool comparison and reporting} 
Each SonarQube-reported vulnerability was manually verified for validity and cross-checked against CodeQL at the same file and location, after which per-language and per-model tables and matrices were compiled.\\
\BfPara{QC} discrepancies documented in tables, and supplementary spreadsheets in the replication package (Sec.~\ref{sec:AnalysisResults}, Sec.~\ref{sec:Conclusion}).};

\draw[->, line width=1pt, black!70] (step1) -- (step2);
\draw[->, line width=1pt, black!70] (step2) -- (step3);
\draw[->, line width=1pt, black!70] (step3) -- (step4);
\draw[->, line width=1pt, black!70] (step4) -- (step5);
\draw[->, line width=1pt, black!70] (step5) -- (step6);
\draw[->, line width=1pt, black!70] (step6) -- (step7);

\end{tikzpicture}
\caption{End-to-end quality-control, covering task design, code generation, compilation, semantic testing and scoring, static analysis, cross-tool comparison, and reporting.}
\label{fig:qc-pipeline}
\end{figure}

\section{Dataset}\label{sec:Dataset_Description}
\rev{To advance the evaluation of the quality and security of LLM code generation, we developed a dataset that reflects practical and security-relevant challenges in real-world programming domains. Compared to previous work, the dataset offers several advantages: it is created manually, language-independent, and tailored to evaluate LLM-generated code across multiple dimensions, including code correctness, maintainability, reliability, and security. Our goal is to provide the research community with a robust foundation for testing, comparing, and improving LLMs in terms of functional correctness, code quality, and security-oriented scenarios by integrating tasks rooted in known vulnerabilities (e.g., CWEs) and actual programming practices, thereby enhancing their overall utility.}

\rev{What distinguishes our dataset is both its breadth, spanning 200 tasks across diverse categories, and its depth, with each task annotated using expert-reviewed tags and accompanied by unit tests designed to evaluate LLM-generated code. The dataset is model-agnostic, supports four programming languages, and adheres to secure coding best practices. Thus, it functions as benchmarking resource and a flexible training and diagnostic tool for researchers and practitioners engaged in LLM-assisted code development.}

\subsection{Dataset Construction} 
\label{sec:dataset_Construction}
\rev{Our dataset construction methodology consists of four major steps: functional categorization of tasks, task sources and cross-language adaptation, task validation and peer review, and task tagging and redundancy reduction. In the following, we review those steps in detail. }

\BfPara{Functional Categorization} We first organized the tasks into functional categories, reflecting both foundational programming skills and real-world software security concerns: 
\begin{enumerate}[leftmargin=*]
    \item[\ding{172}] {\em Secure Coding}: tasks grounded in known software weaknesses based on MITRE's CWE taxonomy, including issues such as injection flaws, cryptographic misuse, and insecure data handling; 
    \item[\ding{173}] {\em Data Structures and Algorithms}: problems focusing on core computing concepts such as array manipulation, recursion, sorting, searching, and graph traversal; 
    \item[\ding{174}] {\em Parsing and Validation}: tasks requiring careful input processing, constraint enforcement, and boundary checking to assess robustness and correctness; 
    \item[\ding{175}] {\em Networking}: scenarios that emulate networked systems, including authentication errors and protocol misuse; 
    \item[\ding{176}] {\em Mathematics and Logic}: challenges designed to evaluate numerical reasoning, precision, and logical flow; 
    \item[\ding{177}] {\em Programming Systems and Utilities}: utility tasks involving file handling, scripting, system configuration, and general-purpose development workflows; 
    \item[\ding{178}] {\em Concurrency and Synchronization}: tasks targeting race conditions, thread safety, and synchronization issues in parallel execution contexts.
\end{enumerate}

\BfPara{Task Sources and Cross-Language Adaptation}
We sourced tasks from coding challenge platforms, including LeetCode~\cite{leetcode}, Edabit~\cite{Edabit}, and Codewars~\cite{Codewars}, as well as from vulnerability databases~\cite{CWEList4_6} and academic datasets~\cite{SiddiqS22}. When suitable tasks were unavailable, we designed additional tasks based on realistic software engineering scenarios. We formulated each task in neutral language to support consistent implementation in Python, Java, C++, and C, enabling fair cross-language comparison. Tasks originally defined for a single language, such as Python tasks from SecurityEval, were rewritten for cross-language applicability without changing core logic, intent, or security focus.

Our design choice of concise, self-contained programming tasks, adopted partially from popular coding challenge platforms, is consistent with recent large-scale evaluations of LLM-generated code on tasks from such sources, where problems with clear specifications and automated tests are used to enable systematic comparison across models~\cite{WangSDUTSZSQ25,MerkelD25,XiaSWLSWHX25,RaihanGP25}. In contrast to these works, our benchmark additionally incorporates security-focused tasks and supports four programming languages, while preserving the same style of well-specified, unit-testable problems. Table~\ref{Table:Distribution_of_tasks_by_source} summarizes the distribution of tasks by source and categorizes them by difficulty level. Only a small subset of the 200 is labeled \emph{Easy}, while the large majority fall into the \emph{Medium} or \emph{Hard} categories. This distribution is intentional and reflects our focus on non-trivial problems that require reasoning about real-world cases, input validation, and security-relevant behavior rather than purely direct tasks.

\BfPara{Task Validation and Peer Review Process} \rev{To ensure the quality, clarity, and reliability of each task in the dataset, we implemented a structured peer review process involving two (professional) software engineers working alongside the second author of this study. The reviewers systematically assessed each task for clarity, difficulty level, and solvability. Their feedback was discussed and incorporated through multiple review cycles, with any ambiguities, inconsistencies, or edge cases resolved collaboratively.
}

\rev{
Unit test is used to verify task correctness, where we established a unified unit test specification applied consistently across all studied programming languages. This specification defined identical inputs, expected outputs, and coverage scenarios, with only language-specific syntax that varied between implementations. An illustrative example of these test scenarios is provided in Appendix~E.  
This unified method ensures that variations in results are attributable to differences in code generation quality rather than inconsistencies in validation, thus providing a fair and controlled basis for comparing LLM-generated codes across languages.}

\begin{table}[t]
\centering
\caption{\normalfont Task distribution by source and difficulty: \underline{E}asy, \underline{M}edium, and \underline{H}ard.}
\label{Table:Distribution_of_tasks_by_source}\vspace{-2mm}
\scalebox{0.85}{\begin{tabular}{p{2.1cm}p{4.0cm}|ccc|l}
\xl{2}
\multirow{2}{*}{\textbf{Source}}  
& \multirow{2}{*}{\textbf{Description}}
& \multicolumn{3}{c|}{\textbf{Difficulty}}
& \multirow{2}{*}{\textbf{Total}}\\
\cline{3-5}
& & \textbf{E} & \textbf{M} & \textbf{H} & \\ \xl{1}
Manual  & Custom tasks written by the first author of this research & 3 & 117 &  3 & 123\\
Code challenge website  & Adopted from Codewars~\cite{Codewars}, LeetCode~\cite{leetcode}, Exercism~\cite{Exercism}, and Edabit~\cite{Edabit} focused on algorithms, data structures and system design & 14 & 43 &  7& 64\\
Security dataset  & Adopted from SecurityEval~\cite{SiddiqS22} and CodeQL ~\cite{CodeQLCW82} focused on code security & 0 & 13 &  0& 13\\\xl{2}
\end{tabular}}
\vspace{-4mm}
\end{table}

\BfPara{Task Tagging and Redundancy Reduction} \rev{Each task was manually reviewed and tagged by at least two reviewers according to its functionality and any implicit security objectives. Tasks covering multiple concepts were assigned multiple tags. This approach enables the evaluation of LLM performance on both narrowly focused and multi-dimensional tasks. To avoid semantic redundancy, we conducted a manual similarity analysis across task prompts and implementations, removing tasks that exhibited significant overlap in logic or structure. In future work, we plan to incorporate automated semantic similarity techniques to further validate dataset diversity and reduce potential bias.}


\subsection{Dataset Features and Improvements} The dataset consists of several attributes to guide the evaluation, including: \ding{172} task number, which refers to unique identifier assigned to each question; \ding{173} prompt title, which refers to brief title that summarizes the problem statement; \ding{174} description, a detailed description of the task, outlining the problem to be solved; \ding{175} hints, which refers to instructions that provided to AI model to guide the generation of the code; \ding{176} solutions, represent the code generated by the LLM for each programming language along with the name of the model used; \ding{177} source, which refers to the origin of the task, whether manually created or adapted; \ding{178} test cases, which refers to 10 test cases per prompt were written to evaluate the semantic of the generated code; \ding{179} tags; labels assigned to each task to facilitate filtrating, categorization, and statistical analysis; \ding{180} comments, which refers to notes made by reviewers to document the evaluation process, including issues encountered during the generation process.
   
\rev{To improve interpretability and enable more focused evaluation, we consolidated the original set of task tags into a structured set of seven semantic categories. Tags were initially assigned through manual annotation based on detailed task analysis; however, as the dataset expanded, the tag set exhibited redundancy, inconsistent phrasing, and varying levels of specificity.} \rev{To address this, we applied a {\em semantic grouping} strategy by clustering related tags under broader functional themes. For instance, technical concepts such as \textit{arrays}, \textit{sorting}, \textit{graphs}, and \textit{recursion} were unified under \textbf{Data Structures and Algorithms}, reflecting core computational topics. Likewise, all tags referencing Common Weakness Enumeration (CWE) identifiers, including \textit{CWE-20}, \textit{CWE-89}, and \textit{CWE-787}, were categorized under \textbf{Secure Coding (CWEs)}, encompassing tasks focused on recognizing and addressing software vulnerabilities. Table~\ref{tab:tag_distribution_grouped} shows how tasks are distributed according to tag categories, achieved by combining similar tags into larger functional groups. A task may belong to one or several categories.}

To illustrate dataset structure and content, Appendix~B presents representative tasks from each programming category. These examples reflect the range of real-world scenarios in the benchmark and contextualize the evaluation by showing the scope and complexity of the prompts. We release the complete dataset as part of the replication package, including task descriptions, difficulty labels, unit tests, and model outputs, enabling detailed inspection and reproduction or extension of the evaluation.

\begin{table}[t]
\centering
\caption{\normalfont Distribution of tasks by tag categories after grouping related tags into broader functional areas. One task might fall into one or multiple categories.}
\label{tab:tag_distribution_grouped} \vspace{-3mm}
\scalebox{0.96}{
\begin{tabular}{lc}
\hline
\textbf{Category} & \textbf{Task Count} \\
\hline
Secure Coding (CWEs - All CWE identifiers grouped) &102 \\
Data Structures and Algorithms & 86 \\
Parsing and Validation & 54 \\
Networking & 48 \\
Mathematics and Logic & 26 \\
Programming Systems and Utilities & 18 \\
Concurrency and Synchronization & 6 \\
\bottomrule
\end{tabular}
}\vspace{-3mm}
\captionsetup{labelfont={color=red}, textfont={color=red}}
\end{table}

\subsection{Secure Coding Coverage via CWE Grouping} 
\rev{To assess model behavior on security-critical tasks, we integrated 86 unique CWE identifiers into our dataset, as noted in the task tagging. Rather than presenting these individually, we grouped them into higher-level vulnerability domains based on their semantic alignment and the MITRE taxonomy. These categories include Input Validation (\ie, CWE-20, CWE-707), Injection (\ie, CWE-89, CWE-78), Buffer Errors (\ie, CWE-119, CWE-787), Access Control Issues, Cryptographic Issues, and others. This organization provides a clearer view of the dataset's security coverage and enables targeted evaluation of model resilience against specific classes of vulnerabilities. Table~\ref{tab:cwe_summary} presents the distribution of tasks across these security domains, and the full list of CWEs is available in our project repository.}

\begin{table}[t]
\centering
\caption{\normalfont MITRE-aligned categorization of representative CWEs in the dataset.}
\label{tab:cwe_summary}\vspace{-3mm}
\scalebox{0.99}{
\begin{tabular}{ll}
\hline
\textbf{CWE Category} & \textbf{Representative CWEs} \\
\hline
Input Validation & CWE-20, CWE-707, CWE-642 \\
Injection & CWE-89, CWE-78, CWE-74 \\
Buffer Errors & CWE-119, CWE-787, CWE-125 \\
Access Control Issues & CWE-285, CWE-639, CWE-287 \\
Cryptographic Issues & CWE-321, CWE-326, CWE-327 \\
Information Exposure & CWE-209, CWE-532 \\
Resource Management Errors & CWE-400, CWE-770 \\
Memory Errors & CWE-416 \\
\hline
\end{tabular}
}\vspace{-3mm}
\end{table}

\section{Results and Discussion}\label{sec:AnalysisResults}

We analyze the performance of different LLMs with respect to code generation, including correctness, security, and reliability. The evaluation process depends on different metrics, \ie compilation-time errors, security, and overall accuracy across different languages. Each LLM's output is compared to define the best generation tool. 

\subsection{Semantic Soundness}
\BfPara{Compilation-Time Errors}
The first part of \autoref{tab:compilation_error} presents insights into the performance of different LLMs in generating compilable code without errors in four programming languages (Java, Python, C++, and C). To calculate the compilation success rate, let \( P \) be the programming language,  \( C \) be the number of compilable files in \( P \), \( T \) be the total number of files from in \( P \), and \( S_P \) be the compilation success rate (in percentage). We then define $C = \sum_{i=1}^{n} C_i$, where \( C_i \) is \( 1 \) if file \( i \) is compileable successfully and \( 0 \) otherwise. Moreover, we define $T = n$, where \( n \) is the total number of files. We also define the compilation success rate in \( P \) as $S_P = ( {C_P}/{T_P} ) \times 100$.

The findings reveal that all models achieve high percentages of compilable code in Python, using three LLMs ({\em gemini-1.5}, {\em codestral}, and {\em GPT-4o}), achieving a success rate 100\%. This suggests that Python's simpler syntax and dynamic nature make it easier for the AI tool to generate accurate and error-free code.  

\begin{table}
\centering
\caption{\normalfont Breakdown (\%) of LLM-generated code files without compilation-time errors (error free) and semantic issues.}
\label{tab:compilation_error}\vspace{-3mm}
\scalebox{0.81}{
\begin{tabular}{lrrrr}
\xl{2}
\multirow{2}{*}{Model} & \multicolumn{4}{c}{{Compilation-time error-free}}  \\ 
\cline{2-5}
& {Java}  & {Python}   & {C++}  & {C}    \\ 
\xl{1}

{\em claude-3.5}& 95.0 & 99.5 & 81.5 & 96.0 \\
{\em gemini-1.5}& 88.5 & 100.0 & 77.5 & 90.5 \\ 
{\em codestral} & 88.5 & 100.0 & 80.0 & 91.5 \\ 
{\em GPT-4o} & 94.0 & 100.0 & 89.0 & 91.5 \\ 
{\em llama-3}& 88.0 & 100.0 & 77.0 & 88.0 \\ 
\xl{2}
\end{tabular}}~\label{tab:generated_code_semantic}
\scalebox{0.81}{
\begin{tabular}{rrrr}
\xl{2}
\multicolumn{4}{c}{{Semantic error-free}}  \\ 
\cline{1-4}
 {Java}  & {Python}   & {C++}  & {C}    \\ 
\xl{1}
91.8 &	92.1 &	91.2 &	84.2 \\
89.5 &	90.5 &	81.0 &	77.0 \\
82.9 &	88.6 &	83.7 &	73.7 \\
88.3 &	90.9 &	88.0 &	81.1 \\
81.6 &	88.5 &	79.3 &	72.9 \\
    
\xl{2}
\end{tabular}
}
\vspace{-3mm}
\end{table}

\addvspace{3pt}
\noindent\fbox{\begin{minipage}{26em}
\BfPara{Takeaway} Python had the highest success rates for compilable code across all models. This suggests that Python's simpler syntax and dynamic nature contribute to the accuracy and reliability of code generation by AI models.
\end{minipage}}

The variability in compilation-time error-free rates reported in \autoref{tab:compilation_error} for Java, C++, and C indicates that LLMs exhibit different strengths depending on the programming language. For Java, performance fluctuates more than in Python, with success rates ranging from 88.00\% to 95.00\%. Models such as {\em claude-3.5} and {\em GPT-4o} outperform others, while {\em llama-3}, {\em gemini-1.5}, and {\em codestral} show slightly lower rates at 88\%, 88.5\%, and 88.5\%, respectively.

These results can be attributed to Java's verbose, statically typed nature, which reduces compilation-time errors when an LLM correctly interprets its syntax and principles. The slight performance differences likely reflect how effectively each LLM handles Java's syntax rules, exception handling, and object-oriented features (\ie encapsulation). Models such as {\em claude-3.5} and {\em GPT-4o} appear to be better fine-tuned for Java-specific dependencies and required imports. Missing import statements, such as \texttt{java.util.*}, remain a common cause of compilation failures.

\addvspace{3pt}
\noindent\fbox{\begin{minipage}{26em}
\BfPara{Takeaway} Models like {\em claude-3.5} and {\em GPT-4o} perform better on Java due to improved handling of syntax rules and package dependencies, while {\em llama-3}, {\em gemini-1.5}, and {\em codestral} show lower success rates, underscoring the importance of Java-specific fine-tuning to improve performance.
\end{minipage}}

C++ shows the lowest success rates across all models, with scores ranging from 77.00\% to 89.00\%. The worst error-free score in C++ is with {\em llama-3} and {\em gemini-1.5}. The performance of LLMs in generating C++ code appears to be constrained by missing include statements, incorrect type handling, and misinterpretation of APIs. C++ is a complex language with many different standards (\ie C++11, C++17), and LLMs struggle when dealing with the complexities of the language's syntax, library usage, and type system, leading to frequent compile-time errors when generating more advanced code involving libraries like STL (Standard Template Library) or 3rd party APIs/libraries.

\addvspace{3pt}
\noindent\fbox{\begin{minipage}{26em}
\BfPara{Takeaway} LLMs struggle with C++ due to missing includes, type errors, standard mismatches (\ie C++11 vs.\ C++17), and API misinterpretations. The language's syntax complexity and strict type system lead to frequent compile-time errors, especially with STL and third-party libraries.
\end{minipage}}

The varying success rates in C code generation reflect differences in handling C standards and dependencies. The high success rate of {\em claude-3.5} (96\%) indicates adherence to C conventions, including proper header inclusion and type declarations. In contrast, {\em llama-3}, with the lowest success rate (88.00\%), frequently produced errors involving undeclared types (\ie \texttt{bool}), missing headers (\eg \texttt{cgi/cgi.h}), and incorrect function arguments. These issues suggest reliance on non-standard libraries or flawed assumptions about the compilation environment. Such inconsistencies highlight the need to generate code that follows standard C practices, includes necessary dependencies, and respects function signature conventions to improve compilation.

\addvspace{3pt}
\noindent\fbox{\begin{minipage}{26em}
\BfPara{Takeaway} The frequent errors in C are related to unknown types (\ie bool), missing headers (\ie \texttt{cgi/cgi.h}), and incorrect function arguments, are due to reliance on non-standard libraries. Ensuring generated code aligns with standard C practices and includes all dependencies with its source is crucial for successful compilation.
\end{minipage}}

The percentage of code files without compilation errors provides insight into how well these AI tools are tuned for different programming paradigms and syntax complexities. Regarding {\em claude-3.5}, it performs well in Java with 95\% as a result, and in C with 96\%, showing its good performance with both object-oriented and functional programming languages. Less performance with the C++ programming language with 81.5\%. {\em GPT-4o} achieved strong results in all program languages, especially C++ (89\%), showing that it is reliable and adaptable for generating error-free code. Regarding {\em gemini-1.5}, this model achieved the highest performance in Python and good in C, but this model achieved the lowest performance as results regarding Java and C++ compared to {\em claude-3.5} and {\em GPT-4o}. Lastly, regarding {\em llama-3}, performs lower in most languages except Python. It struggles with Java (88\%), C++ (77\%), and C (88\%), showing that it is not as well tuned for more complex languages. The reason behind the results shown for each of the AI tools refers to a set of factors such as training data, since each model trained on large, more diverse, and higher-quality datasets is better at understanding various programming languages \cite{brown2020language}. Another important factor regarding the performance of the AI tool refers to model size and complexity, since a larger model with more parameters such as {\em GPT-4o} can better understand and generate.

\BfPara{Types of Compilation-Time Errors} 
A breakdown of the compilation-time errors is summarized in \autoref{tab:compilation_time_errors_java}. Python's breakdown is excluded from the table since it lists only error types occurring more than once across all LLMs. In the following, compilation errors are grouped into categories to understand the common failure modes across the models. 

\begin{enumerate}[leftmargin=*] 
\item[\ding{172}] {\em Library Errors.}
The most common error category across all models involved missing imports or incorrect package references. Frequent issues included ``cannot find symbol'' errors and failures to recognize or include required third-party libraries, particularly missing imports such as \texttt{java.util.*}, especially in tasks involving non-standard libraries. These patterns suggest a need for better handling of Java libraries and packages, potentially through training LLMs on more diverse codebases with correct import usage. In addition, improving the contextual understanding of external dependencies may enhance model performance in such tasks.

\item[\ding{173}] {\em Exception Handling.}
Missing or incorrect exception handling was another area where the models faltered. {\em gemini-1.5}, {\em codestral}, and {\em llama-3} were particularly prone to this type of error, indicating that these models could improve by focusing on the need to handle exceptions based on the exception that can be triggered by code in scope for a particular code block or method.
\item[\ding{174}] {\em Missing Class/Member.}
An often recurrent issue flagged by {\em codestral} is the absence of a class or class member, such as setter and/or getter methods. This error arises when the model presumes the presence of a specific method or class without including the pertinent code or even suggesting it within comments.
\item[\ding{175}] {\em Syntax}
Syntax errors, including incorrect symbols, missing semicolons, or invalid syntax, were found primarily in {\em codestral}. This pattern suggests a need for improved fine-tuning on Java syntax conventions and a deeper understanding of language-specific rules.

\item[\ding{176}] {\em Type Compatibility.}
Errors related to incompatible types, such as attempts to assign a java.lang.Object to a java.lang.String, or java.io.IOException cannot be converted to java.lang.SecurityException. These errors indicate the models' ability to infer the data types in context.

\item[\ding{177}] {\em Variable Identifier.}
Errors arising from the use of undefined variables were found in multiple models. This indicates that the models occasionally fail to maintain variable state consistency across different sections of the code.
\item[\ding{178}] {\em Undeclared Var./Fun.}
Errors where a variable or function is used before being declared, leading to a failure in recognizing identifiers. This may happen if a necessary declaration or definition is missing.

\item[\ding{179}] {\em Binding/Qualifier.}
Errors related to failing to bind a variable correctly or mismatches with qualifiers. These errors often occur in function parameters or references.

\item[\ding{180}] {\em Constant Undeclared.}
 Errors with undeclared constants like {\tt true} and {\tt nullptr}.
\end{enumerate}

\begin{table*}
\centering
\caption{{\normalfont Compilation-time error types in generated code. LLMs: \ding{182} {\em claude-3.5}, \ding{183} {\em gemini-1.5}, \ding{184} {\em codestral}, \ding{185} {\em GPT-4o}, \ding{186} {\em llama-3}.}}
\label{tab:compilation_time_errors_java}\vspace{-3mm}
\scalebox{0.78}{
\begin{tabular}{lccccc}
\xl{2}
\multirow{2}{*}{Error Type} & \multicolumn{5}{c}{{Java}}  \\ 
\cline{2-6}
  & \ding{182}   & \ding{183}  & \ding{184} & \ding{185} & \ding{186}  \\ 
\xl{1}
{\em Library Errors}  & 10 & 17 & 15 & 8&17 \\ 
{\em Exception Handling } & 0 & 3 & 2 & 0&4 \\
{\em Class/Member}  & 0 & 0 & 4 & 0&1 \\
{\em Syntax}& 0& 1&1 & 2 & 0 \\
{\em Type Compatibility}  & 0 & 1 & 1 & 0& 2 \\ 
{\em Variable def. related}  & 0 & 1 & 1 & 1& 1 \\

\xl{2}
\end{tabular}}~
\label{tab:compilation_time_errors_cpp}
\scalebox{0.78}{
\begin{tabular}{lccccc}
\xl{2}
\multirow{2}{*}{Error Type} & \multicolumn{5}{c}{{C++}}  \\ 
\cline{2-6}
  & \ding{182}   & \ding{183}  & \ding{184} & \ding{185} & \ding{186}  \\ 
\xl{1}
{\em Library Errors} & 20 & 8 & 4 & 11 & 22 \\ 
{\em Binding/Qualifier} & 0 & 1 & 0 & 0 & 2 \\ 
{\em Class/Member} & 3 & 2 & 9 & 3 & 0 \\ 
{\em Syntax } & 4 & 2 & 6 & 1 & 6 \\ 
{\em Variable def. related } & 6 & 15 & 5 & 4 & 4 \\ 
{\em Undeclared Var./Fun. } & 2 & 13 & 11 & 2 & 6 \\

\xl{2}
\end{tabular}}
\label{tab:compilation_time_errors_c}
\scalebox{0.78}{
\begin{tabular}{lccccc}
\xl{2}
\multirow{2}{*}{Error Type} & \multicolumn{5}{c}{{C}}  \\ 
\cline{2-6}
  & \ding{182}   & \ding{183}  & \ding{184} & \ding{185} & \ding{186}  \\ 
\xl{1}
{\em Library Error} & 7 & 14 & 3 & 3 & 5 \\ 

{\em Constant Undeclared} & 0 & 1 & 2 & 0 & 10 \\ 
{\em Function Argument} & 2 & 0 & 0 & 0 & 3 \\ 
{\em Undeclared Var./Fun.} & 1 & 1 & 4 & 2 & 6 \\

{\em Incomplete Code} & 1 & 1 & 1 & 1 & 0 \\ 
 {\em Other } & 2 & 3 & 7 & 1 & 0 \\ 

\xl{2}
\end{tabular}}\vspace{-3mm}
\end{table*}

\BfPara{\rev{Semantic Correctness Using Unit Test}} \rev{The results in \autoref{tab:generated_code_semantic} summarize semantic correctness across languages. Most models exhibit consistent performance, though accuracy varies with language complexity. {\em claude-3.5} achieved up to 92.1\%, while {\em GPT-4o} ranged from 90.9\% (Python) to 81.1\% (C). {\em gemini-1.5} showed a broader spread, from 90.5\% to 77.0\%, and {\em codestral} ranged from 88.6\% to 73.7\%. {\em llama-3} yielded the lowest scores overall, peaking at 88.5\% in Python. These differences underscore the sensitivity of LLM outputs to both prompt formulation and target language. Although tasks require similar reasoning, performance discrepancies suggest underlying differences in how models interpret and implement logic, likely influenced by training data, architecture, or optimization strategies. The wider variability seen in {\em gemini-1.5}, {\em llama-3}, and {\em codestral} may reflect challenges in handling specific language features or complex constructs, pointing to design or training gaps.}

\addvspace{3pt}
\noindent\fbox{\begin{minipage}{26em}
\BfPara{Takeaway} The variance in success rates \rev{between different LLMs, with an average of 14.7\% for {\em gemini-1.5}, {\em llama-3}, and {\em codestral}}, indicates notable differences in how these models handle coding tasks, likely due to variations in training data, model design, and optimization strategies.
\end{minipage}}

For Java, the models generally showed strong performance, with {\em claude-3.5} achieving the highest success rate of \rev{91.8\%}. This indicates a good proficiency in creating effective and precise Java code, probably because of Java's object-oriented design and consistent libraries, which apparently correspond well with the models' training data. {\em gemini-1.5} also performed well, reaching a success rate of \rev{89.5\%}, demonstrating its proficiency in managing Java syntax and standard programming constructs. {\em codestral} and \rev{{\em llama-3}} showed slightly lower success rates at \rev{82.9\%} and \rev{81.6\%}, respectively. The variation in performance, especially for {\em codestral} \rev{and {\em llama-3}}, could indicate difficulties in addressing Java's more complex class structures or standard library usage, which aligns with its relatively lower compilation success rate. On the whole, the high semantic correctness across models confirms their capacity to generate functional Java code, while also highlighting areas where deeper reasoning or structural precision remains necessary.

\addvspace{3pt}
\noindent\fbox{\begin{minipage}{26em}
\BfPara{Takeaway} 
LLMs perform strongly on Java, with {\em claude-3.5} achieving the highest score at \rev{91.8\%}, likely reflecting Java's structured object-oriented design and stable libraries. {\em gemini-1.5} follows at \rev{89.5\%}, while {\em codestral} and \rev{{\em llama-3}} score lower at \rev{82.9\%} and \rev{81.6\%}, respectively. These results indicate overall proficiency but also expose remaining challenges in handling Java-specific complexity.
\end{minipage}}

Among the evaluated languages, Python achieved the highest success rates, with {\em claude-3.5} leading at \rev{92.1\%}, followed by {\em GPT-4o} and {\em gemini-1.5} at \rev{90.9\%} and \rev{90.5\%}, respectively. Python's simplicity and dynamic typing likely contribute to these high success rates, as its syntax is generally more permissive and flexible than statically typed languages. The consistently strong performance across models suggests they are well-versed in Python constructs. Notably, {\em llama-3} achieved a slightly lower success rate of \rev{88.5\%}, indicating that even high-performing models may encounter challenges in generating correct Python code, potentially due to differences in language handling or reasoning strategies. For C++, {\em claude-3.5} achieved the highest semantic success rate at 91.2\%, demonstrating a strong capability in handling C++ syntax and object-oriented constructs. {\em GPT-4o} followed at 88.0\%, showing reliable performance on complex language features. {\em codestral} reached 83.7\%, while {\em gemini-1.5} and {\em llama-3} trailed at 81.0\% and 79.3\%, respectively, indicating more pronounced challenges in generating correct C++ code across these models. Although these trends, such as Python demonstrating fewer compilation errors than C++, align with practitioners' expectations, our results quantify the magnitude of these differences across models and link them to specific language quality and security outcomes.

\addvspace{3pt}
\noindent\fbox{\begin{minipage}{26em}
\BfPara{Takeaway} 
{\em claude-3.5} achieves the highest success rate in C++ at \rev{91.2\%}, then {\em GPT-4o} at \rev{88.0\%} and {\em codestral} at \rev{83.7\%}, indicating differences in handling C++ complexity. C remains more challenging across models due to stricter syntax, with {\em claude-3.5} leading at \rev{84.2\%} and {\em llama-3} scoring lowest at \rev{72.9\%}. These results underscore ongoing limitations in low-level language reasoning and generation.
\end{minipage}}

Success rates in C were notably lower across all models, with {\em claude-3.5} achieving the highest at \rev{84.2\%}, followed by {\em GPT-4o} at \rev{81.1\%}, {\em gemini-1.5} at \rev{77.0\%}, {\em codestral} at \rev{73.7\%}, and {\em llama-3} at \rev{72.9\%}. This performance gap can be attributed to C's stricter requirements, including manual memory management, explicit type declarations, and the mandatory inclusion of standard headers. These findings underscore the need for targeted enhancements in LLM-based code generation to better address the syntactic rigor and low-level control demanded by the C programming language.

\addvspace{3pt}
\noindent\fbox{\begin{minipage}{26em}
\BfPara{Takeaway}
Semantic correctness is high in Python and Java for top-performing models, but drops in C++ and especially C, where success rates range roughly from 73\% to 84\%. This gap shows that the benchmark is not uniformly easy and exposes language-dependent challenges, particularly in lower-level languages with manual memory management and stricter type systems.
\end{minipage}}

\rev{\subsection{Complexity}}
This section evaluates the static code features of LLMs generated code across Java, Python, C++, and C using several complexity metrics, such as lines of code (LoC), cyclic complexity (CyC), cyclic complexity density (CCD) and cognitive complexity (CoC). The analysis included code generated by various LLMs identified as {\em claude-3.5}, {\em gemini-1.5}, {\em codestral}, {\em GPT-4o}, and {\em llama-3}. The results are provided in~\autoref{tab:static_features_java}. For Java code, as summarized in~\autoref{tab:static_features_java}, shows that {\em claude-3.5} produced the highest LoC (6,480) and the highest CoC (891). {\em gemini-1.5} showed the highest CCD (0.23), indicating higher complexity. Conversely, {\em codestral} generated relatively less complex code with the lowest CyC (730) and CoC (585). The evaluation of the Python code showed that among the models, {\em gemini-1.5} had the highest CCD (0.37), reflecting a very dense complexity despite having a lower LoC (2,077). {\em claude-3.5} produced the highest LoC (3,474) and the highest CoC (881) the same as {\em gemini-1.5}. {\em codestral} presented the lowest values in all metrics, indicating simpler and less complex code. 

For the C++ code, the results indicate that {\em claude-3.5} generated the highest LoC (6,725) and the highest CoC (1,096). The highest CCD was observed with {\em gemini-1.5} (0.25), suggesting more intricate code patterns. In contrast, {\em GPT-4o} showed the lowest CyC (676) and CoC (510), indicating simpler, more maintainable code.

With respect to the C code, {\em claude-3.5} again led with the highest LoC (7,509) and CoC (1,463). {\em gemini-1.5} had the highest CCD (0.28), while {\em codestral} presented with the least CyC (894) and CoC (869), indicating a simpler code.

Based on the analysis above, the CCD varies between languages, reflecting differences in code density and distribution. CoC closely follows CyC but offers additional insight into code readability and maintainability.~{\em claude-3.5} generally produces more verbose and complex code across all languages, suggesting greater difficulty in readability and even the runtime overhead. Lastly, {\em codestral} and {\em GPT-4o} tend to generate simpler and more maintainable code, with lower complexity metrics.

\begin{table*}
\centering
\caption{\normalfont Static features: lines of code (LoC), cyclomatic complexity (CyC), density (CCD), and cognitive complexity (CoC).}
\label{tab:static_features_java}\vspace{-3mm}
\scalebox{0.85}{
\begin{tabular}{lcccc}
\xl{2}
\multirow{2}{*}{Model} & \multicolumn{4}{c}{{Java}}  \\ 
\cline{2-5}
& {LoC}  & {CyC}   & {CCD} & {CoC}   \\ 
\xl{1}
{\em claude-3.5} & 6,480 & 1,097& 0.17 & 891\\ 
{\em gemini-1.5} & 5,606 & 965 & 0.23 & 902\\
{\em codestral} & 4,143 & 730 & 0.13  & 585 \\
{\em GPT-4o}    & 5,328 & 908 & 0.17 & 670 \\
{\em llama-3}   & 4,993 & 913 & 0.18 & 809\\ \xl{1}
        \textbf{Summary} & \textbf{26,550} & \textbf{4,613} & \textbf{0.17} & \textbf{3,857}\\ 

\xl{2}
\end{tabular}}
\label{tab:static_features_python}
\scalebox{0.85}{
\begin{tabular}{cccc}
\xl{2}
\multicolumn{4}{c}{{Python}}  \\ 
\cline{1-4}
{LoC}  & {CyC}   & {CCD} & {CoC}  \\ 
\xl{1}
3,474 & 895 & 0.26 & 881\\ 
3,064 & 765 & 0.37 & 881\\ 
2,077 & 553 & 0.18 & 491\\ 
2,906 & 665 & 0.23 & 610\\ 
2,633 & 677 & 0.26 & 631\\ \xl{1}
 \textbf{14,154} & \textbf{3,555} & \textbf{0.25} & \textbf{3,494} \\ 

\xl{2}
\end{tabular}}~
\label{tab:static_features_cpp}
\scalebox{0.85}{
\begin{tabular}{cccc}
\xl{2}
\multicolumn{4}{c}{{C++}}  \\ 
\cline{1-4}
{LoC}  & {CyC}   & {CCD} & {CoC}  \\ 
\xl{1}

6,725 & 1,261 &  0.19 & 1,096 \\ 
5,822 & 1,084 & 0.25 & 1,044 \\
4,293 & 825  &  0.14 & 732 \\
5,263 & 676  & 0.13  & 510 \\ 
5,077 & 979  &  0.19  &  884 \\ \xl{1}
\textbf{27,180} & \textbf{4,825} & \textbf{0.18}& \textbf{4,266}\\
\xl{2}
\end{tabular}}
\label{tab:static_features_c}
\scalebox{0.85}{
\begin{tabular}{lcccc}
\xl{2}
\multicolumn{4}{c}{{C}}  \\ 
\cline{1-4}
{LoC}  & {CyC}   & {CCD} & {CoC}   \\ 
\xl{1}
7,509 & 1,464 & 0.19  & 1,463 \\ 
6,555 & 1,256 &  0.28 & 1,331 \\ 
4,532 & 894   &  0.14 & 869 \\ 
5,926 & 1,133 &  0.19 & 994 \\ 
5,006 & 1,004 & 0.20  & 1,012 \\ \xl{1}
\textbf{29,528} & \textbf{5,751} & \textbf{0.19} &\textbf{5,669} \\    
\xl{2}
\end{tabular}}\vspace{-3mm}
\end{table*}

\rev{\subsection{Security and Quality Attribute}} The security and quality evaluation is based on the different metrics in~\autoref{sec:Methodology}, including LoC, the number of lines of code, which affects the complexity, since a large LoC can indicate more functionality, but may also lead to higher maintenance demands \cite{trisovic2022large}. 

\autoref{tab:quality_attribute_java} presents Java results across key quality attributes. For lines of code (LoC), {\em claude-3.5} generates the highest output, contributing to increased complexity and variability. In contrast, {\em codestral} is the most efficient model in terms of LoC. Regarding security, which assesses resistance to unauthorized access and vulnerabilities, {\em gemini-1.5} yields the highest number of vulnerabilities among the LLMs for Java. This outcome is likely due to a relative lack of security-focused training data compared to other models.

\begin{table*}
\centering
\caption{\normalfont Quality attributes against security (S), reliability (R), maintainability (M), and security hotspots (SH).}
\label{tab:quality_attribute_java}\vspace{-3mm}
\scalebox{0.85}{
\begin{tabular}{lccccc}
\xl{2}
\multirow{2}{*}{Model} & \multicolumn{4}{c}{{Java}}  \\ 
\cline{2-5}
  & {S}   & {R}  & {M} & {SH}   \\ 
\xl{1}
{\em claude-3.5}  & 15 & 51 & 810 & 61 \\ 
{\em gemini-1.5 } & 16 & 46 & 694 & 33 \\
{\em codestral}  & 12 & 37 & 535 & 40 \\
{\em GPT-4o}  & 13 & 52 & 925 & 76 \\
{\em llama-3}  & 13 & 46 & 932 & 27 \\ \xl{1}
        \textbf{Summary} & \textbf{69} & \textbf{232} & \textbf{3896} & \textbf{237} \\ 

\xl{2}
\end{tabular}}
\label{tab:quality_attribute_python}
\scalebox{0.85}{
\begin{tabular}{ccccc}
\xl{2}
\multicolumn{4}{c}{{Python}}  \\ 
\cline{1-4}
{S}   & {R}  & {M} & {SH}   \\ 
\xl{1}
 5 & 0 & 82 & 44 \\ 
 9 & 1 & 83 & 36 \\ 
 9 & 1 & 50 & 39 \\ 
 8 & 0 & 75 & 43 \\ 
 9 & 1 & 63 & 34 \\ \xl{1}
\textbf{40} & \textbf{3} & \textbf{353} & \textbf{196} \\ 

\xl{2}
\end{tabular}}~
\label{tab:quality_attribute_cpp}
\scalebox{0.85}{
\begin{tabular}{ccccc}
\xl{2}
\multicolumn{4}{c}{{C++}}  \\ 
\cline{1-4}
{S}   & {R}  & {M} & {SH}   \\ 
\xl{1}

9 & 6 & 590 & 36 \\ 
17 & 8 & 587 & 18 \\
10 & 5 & 470 & 20 \\
7 & 6 & 459 & 17 \\ 
10 & 10 & 565 & 18 \\ \xl{1}
\textbf{53} & \textbf{35} & \textbf{2671} & \textbf{109} \\
\xl{2}
\end{tabular}}
\label{tab:quality_attribute_c}
\scalebox{0.85}{
\begin{tabular}{lccccc}
\xl{2}
\multicolumn{4}{c}{{C}}  \\ 
\cline{1-4}
{S}   & {R}  & {M} & {SH}   \\ 
\xl{1}
19 & 60 & 449 & 186 \\ 
33 & 50 & 352 & 146 \\ 
21 & 39 & 314 & 115 \\ 
27 & 37 & 380 & 133 \\ 
22 & 54 & 322 & 182 \\ \xl{1}
\textbf{122} & \textbf{240} & \textbf{1817} & \textbf{762} \\    
\xl{2}
\end{tabular}}\vspace{-3mm}
\end{table*}

Per~\autoref{tab:quality_attribute_java}, {\em codestral} has the fewest reliability issues (37), indicating minimal crashes or errors, likely due to its lower lines of code compared to other models. It also ranks best in maintainability, with 535 issues, reflecting code that is easier to understand, modify, and extend, attributable to clear structure, comments, and function decomposition. For security hotspots, {\em llama-3} performs best, with only 27 issues.

For the Python code evaluation in~\autoref{tab:quality_attribute_python}, the models show notable variations in quality. {\em claude-3.5}, generating 3,474 lines, has 82 maintainability issues and 44 security hotspots, indicating areas needing security improvements. {\em codestral}, with 2,077 lines, performs better with 50 maintainability issues and 39 security hotspots, reflecting better overall control. {\em gemini-1.5} and {\em GPT-4o}, producing 3,064 and 2,906 lines respectively, show similar outcomes with maintainability issues (83 and 75) and security hotspots (36 and 43). {\em llama-3}, generating 2,633 lines, has the fewest security hotspots (34) and moderate maintainability issues (63), indicating balanced but not flawless performance.

\autoref{tab:quality_attribute_cpp} reveals that {\em claude-3.5}, with 6,725 LoC, has the highest maintainability issues (590) but few security hotspots (36), suggesting strong security but complex maintenance. {\em codestral}, generating 4,293 LoC, has the fewest maintainability issues (470) and security hotspots (20), indicating a balanced performance. {\em gemini-1.5}, producing 5,822 LoC and with a higher count of both security issues (17) and maintainability concerns (587). {\em GPT-4o} and {\em llama-3}, with 5,263 and 5,077 lines, respectively, demonstrate moderate and balanced performance in both areas.

For C, ~\autoref{tab:quality_attribute_c} shows that {\em claude-3.5}, with 7,509 LoC, has a high number of maintainability issues (449) and security hotspots (186), indicating difficulties in managing large and secure code. {\em codestral}, producing 4,532 lines, has fewer security hotspots (115) and maintainability issues (314), though it scores lower in reliability (39). {\em gemini-1.5}, with 6,555 lines, shows the most security issues (33) and moderate maintainability concerns (352), signaling significant security problems. {\em GPT-4o}, generating 5,926 lines, balances moderate security hotspots (133) and maintainability issues (380), while {\em llama-3}, with 5,006 lines, shows higher reliability (54) but faces substantial security hotspots (182).

\rev{
To illustrate our findings, Appendix~C 
presents examples of reliability, maintainability, and security hotspot issues in LLM-generated code. These cases show how seemingly minor flaws such as missing input validation, outdated API usage, or unsafe regular expressions can compromise code quality, even in relatively simple tasks.
}

\BfPara{CWE Categories of Security Quality Attribute} The detection of security flaws in the code produced by five Large Language Models (LLMs) for Java, Python, C, and C++ highlights substantial differences in both the quality and security of the code among the models and programming languages. The \autoref{tab:cwe_errors_java} presents a breakdown of identified CWE categories, reflecting the different LLMs strengths and weaknesses in secure code generation. This analysis sheds light on common security challenges that various LLMs might present when producing code, identifying opportunities to enhance LLM-driven code generation.

In Java, the most observed CWE is CWE-780 (Use of RSA Algorithm without OAEP), which remains consistently high in frequency across various models, notably with models {\em claude-3.5}, {\em gemini-1.5}, and {\em GPT-4o}. This suggests that while generating code for encryption tasks, many LLMs fail to apply the necessary secure padding scheme (OAEP), which is essential for RSA encryption security. This could lead to insecure cryptographic implementations if used in real-world applications. Additionally, CWE-259 (Use of Hard-coded Password) and CWE-295 (Improper Certificate Validation) appear frequently, suggesting that certain models tend to generate credentials or handle certificates in an insecure manner. Issues such as CWE-611 (Improper Restriction of XML External Entity Reference) indicate a common oversight in XML handling, potentially exposing generated code to XML external entity (XXE) attacks.

In Python, CWE-780 is also frequently observed, though less than in Java, indicating that RSA padding issues are not exclusive to Java but persist across languages. Another notable vulnerability in Python is CWE-259, which highlights the common use of hard-coded credentials, posing a risk to sensitive information if deployed. Moreover, Python's handling of CWE-79 (Improper Neutralization of Input During Web Page Generation) reveals weaknesses in web-based output, leading to cross-site scripting (XSS) vulnerabilities.

C++ exhibits a range of vulnerabilities, with CWE-295, CWE-326, and CWE-327 (related to improper certificate validation, inadequate encryption strength, and the use of a broken or risky cryptographic algorithm, respectively) being highly prevalent. These vulnerabilities suggest that cryptographic practices and certificate handling in C++ code generated by LLMs are notably insecure. Additionally, CWE-780 (RSA without OAEP) and CWE-611 (XXE) are also present, indicating potential security weaknesses in cryptographic and XML-handling implementations. This underscores that certain models do not enforce secure practices when generating security-sensitive code in C++. The widespread presence of CWE-297 (Improper Validation of Certificate with Host Mismatch) further highlights weaknesses in certificate validation, which could expose systems to man-in-the-middle (MITM) attack vectors if deployed.

For C, there is a high prevalence of CWE-120 (Buffer Copy without Checking Size of Input), reflecting common buffer overflow vulnerabilities in low-level languages that require manual memory management. The frequent occurrences of CWE-295, CWE-326, and CWE-327 indicate that LLMs often generate C code with poor cryptographic practices and inadequate certificate validation, a trend also observed in C++. Additionally, CWE-131 (Incorrect Calculation of Buffer Size) and CWE-788 (Access of Memory Location After End of Buffer) highlight inadequate buffer management, posing serious security risks such as memory corruption and arbitrary code execution. CWE-780 appears frequently in {\em gemini-1.5}, {\em codestral}, and {\em GPT-4o}, indicating RSA padding issues, though at lower rates than in higher-level languages like Java and Python.

In summary, there is considerable variation in the security quality of the generated code across different programming languages. {\em gemini-1.5} exhibited the highest overall reported vulnerabilities, suggesting less conservative or secure default behaviors. However, some vulnerabilities persist across all models and languages.

Vulnerabilities in languages like C and C++ are more skewed towards memory management issues (\ie buffer overflows, incorrect buffer size calculations) compared to Java and Python, where cryptographic and XML-related vulnerabilities are more common. This highlights inherent language-specific risks, such as memory safety in C and C++ and secure API use in Java and Python.

To illustrate identified security vulnerabilities, Appendix~C presents two representative cases: RSA encryption without OAEP padding (CWE-780) and hard-coded passwords (CWE-259), corresponding to tasks~106 and~83. These examples show that LLMs may default to insecure practices when safer alternatives exist, indicating limited contextual security awareness. These CWEs are well known and documented in prior analyses of LLM-generated and human-written code. The goal here is to quantify how often such known weaknesses, along with other vulnerabilities, appear across recent LLMs and programming languages under a unified benchmark. The patterns in~\autoref{tab:cwe_errors_java} therefore support comparative analysis of model and language behavior rather than claims of novel vulnerability types.

\begin{table}
\centering
\caption{{\normalfont Breakdown of the security quality attributes based on the CWE categories. LLMs: \ding{182} {\em claude-3.5}, \ding{183} {\em gemini-1.5}, \ding{184} {\em codestral}, \ding{185} {\em GPT-4o}, \ding{186} {\em llama-3}.}}
\label{tab:cwe_errors_java}\vspace{-3mm}
\scalebox{0.85}{
\begin{tabular}{lccccc}
\xl{2}
\multirow{2}{*}{CWE ID} & \multicolumn{5}{c}{{Java}}  \\ 
\cline{2-6}
  & \ding{182}   & \ding{183}  & \ding{184} & \ding{185} & \ding{186}  \\ 
\xl{1}
        780 & 6 & 7 & 4 & 8 & 4 \\ 
        502 & 0 & 1 & 1 & 0 & 0 \\ 
        22 & 3 & 3 & 2 & 2 & 3 \\ 
        918 & 0 & 0 & 0 & 0 & 0 \\ 
        259 & 2 & 2 & 2 & 1 & 4 \\ 
        295 & 2 & 0 & 0 & 0 & 0 \\ 
        611 & 2 & 2 & 2 & 2 & 1 \\ 
        79 & 0 & 1 & 1 & 0 & 0 \\ 
        521 & 0 & 0 & 0 & 0 & 1 \\ 
        759 & 0 & 0 & 0 & 0 & 0 \\ 

\xl{2}
\end{tabular}}~
\label{tab:cwe_errors_python}
\scalebox{0.85}{
\begin{tabular}{lccccc}
\xl{2}
\multirow{2}{*}{CWE ID} & \multicolumn{5}{c}{{Python}}  \\ 
\cline{2-6}
  & \ding{182}   & \ding{183}  & \ding{184} & \ding{185} & \ding{186}  \\ 
\xl{1}
        780 & 3 & 4 & 5 & 3 & 3 \\ 
         502 & 0 & 0 & 0 & 0 & 0 \\ 
         22 & 0 & 0 & 0 & 0 & 0 \\ 
        918 & 0 & 1 & 0 & 0 & 0 \\ 
        259 & 2 & 2 & 2 & 2 & 3 \\ 
         295 & 0 & 0 & 0 & 0 & 0 \\ 
         611 & 0 & 0 & 0 & 0 & 0 \\ 
        79 & 0 & 0 & 1 & 1 & 2 \\ 
        521 & 0 & 1 & 1 & 2 & 1 \\ 
        759 & 0 & 1 & 0 & 0 & 0 \\ 
\xl{2}

\end{tabular}}
\vspace{1mm}
\label{tab:cwe_errors_CPP}
\scalebox{0.8}{
\begin{tabular}{lccccc}
\xl{2}
\multirow{2}{*}{CWE ID} & \multicolumn{5}{c}{{C++}}  \\ 
\cline{2-6}
  & \ding{182}   & \ding{183}  & \ding{184} & \ding{185} & \ding{186}  \\ 
\xl{1}

      120 & 0 & 0 & 0 & 0 & 0 \\ 
        295, 326, 327 & 9 & 8 & 6 & 5 & 7 \\ 
        297 & 0 & 2 & 1 & 1 & 2 \\ 
        780 & 0 & 7 & 3 & 1 & 0 \\ 
        611 & 0 & 0 & 0 & 0 & 1 \\ 

\xl{2}
\end{tabular}}
\label{tab:cwe_errors_C}
\scalebox{0.8}{
\begin{tabular}{lccccc}
\xl{2}
\multirow{2}{*}{CWE ID} & \multicolumn{5}{c}{{C}}  \\ 
\cline{2-6}
  & \ding{182}   & \ding{183}  & \ding{184} & \ding{185} & \ding{186}  \\ 
\xl{1}

        120 & 9 & 19 & 14 & 13 & 11 \\ 
        295, 326, 327 & 9 & 5 & 3 & 8 & 7 \\ 
        297 & 1 & 2 & 1 & 2 & 3 \\ 
        131, 788 & 0 & 1 & 0 & 1 & 1 \\ 
        780 & 0 & 6 & 3 & 3 & 0 \\ 

\xl{2}
\end{tabular}}\vspace{-3mm}
\end{table}

\subsection{Clean Code}
\autoref{tab:clean_code_attribute_java} shows the clean code analysis for Java, where most models handled consistency well, with no issues reported except for {\em GPT-4o} (8) and {\em llama-3} (61). However, intentionality issues were prominent, especially for {\em claude-3.5} (203) and {\em gemini-1.5} (244), suggesting that the clarity of code purpose could be improved. Adaptability scores were highest for {\em claude-3.5} (532) and lowest for {\em codestral} (182), indicating varying levels of code flexibility for future changes.

In the same table, the clean code analysis for Python shows slightly more noticeable consistency issues, especially for {\em claude-3.5} (54) and {\em llama-3} (51), while the intentionality attribute remains a common challenge across all models. Adaptability and responsibility issues were low across the board, highlighting Python's inherent simplicity and flexibility. This suggests that while Python code is generally adaptable, models need to improve its clarity.

\begin{table*}
\centering
\caption{\normalfont Evaluation of the clean code in terms of consistency (C), intentionality (I), adaptability (A), and responsibility (R).}
\label{tab:clean_code_attribute_java}\vspace{-3mm}
\scalebox{0.85}{
\begin{tabular}{lccccc}
\xl{2}
\multirow{2}{*}{Model} & \multicolumn{4}{c}{{Java}}  \\ 
\cline{2-5}
& {C}  & {I}   & {A}  & {R}  \\ 
\xl{1}
 
{\em claude-3.5}& 0 & 203 & 532 & 10 \\
{\em gemini-1.5}& 0 & 244 & 387 & 9 \\ 
{\em codestral}& 0 & 181 & 182 & 6 \\ 
{\em GPT-4o} & 8 & 148 & 392 & 9 \\ 
{\em llama-3}& 61 & 200 & 360 & 9 \\ \hline
\textbf{Summary} & \textbf{69} & \textbf{976} & \textbf{1,853} & \textbf{43} \\ 

\xl{2}
\end{tabular}}
\label{tab:clean_code_attribute_python}
\scalebox{0.85}{
\begin{tabular}{cccc}
\xl{2}
\multicolumn{4}{c}{{Python}}  \\ 
\cline{1-4}
{C}  & {I}   & {A}  & {R}  \\ 
\xl{1}
54 & 23 & 5 & 5 \\
47 & 34 & 5 & 7 \\
35 & 17 & 1 & 7 \\
47 & 27 & 4 & 5 \\
51 & 15 & 1 & 6 \\ \xl{1}
\textbf{234} & \textbf{116} & \textbf{16} & \textbf{30} \\ 
\xl{2}
\end{tabular}}
\label{tab:clean_code_attribute_cpp}
\scalebox{0.85}{
\begin{tabular}{cccc}
\xl{2}
\multicolumn{4}{c}{{C++}}  \\ 
\cline{1-4}
{C}  & {I}   & {A}  & {R}  \\ 
\xl{1}
116 & 438 & 42 & 9 \\
136 & 421 & 40 & 15 \\
117 & 335 & 24 & 9 \\ 
93 & 334 & 39 & 6 \\ 
121 & 408 & 49 & 7 \\ \xl{1}
\textbf{583} & \textbf{1,936} & \textbf{194} & \textbf{46} \\ 
\xl{2}
\end{tabular}}
\label{tab:clean_code_attribute_c}
\scalebox{0.85}{
\begin{tabular}{cccc}
\xl{2}
\multicolumn{4}{c}{{C}}  \\ 
\cline{1-4}
{C}  & {I}   & {A}  & {R}  \\ 
\xl{1}
162 & 321 & 36 & 9 \\ 
188 & 197 & 38 & 11 \\
155 & 186 & 27 & 6 \\ 
197 & 210 & 25 & 11 \\
160 & 205 & 25 & 7 \\ \xl{1}
\textbf{862} & \textbf{1,119} & \textbf{151} & \textbf{44} \\ \xl{2}
\end{tabular}}
\end{table*}

With regard to C++, \autoref{tab:clean_code_attribute_c} also shows significant challenges, particularly with intentionality. {\em claude-3.5} (438) and {\em gemini-1.5} (421) had the highest intentionality issues, reflecting difficulties in generating code that clearly communicates its purpose. Consistency was also a challenge for all models, with no significant peaks in reducing the problems. In addition, for the C language, we can see that intentionality and consistency issues are prevalent, with {\em GPT-4o} (197) and {\em gemini-1.5} (188) ranking highest in consistency issues. All models exhibited relatively high adaptability issues, particularly {\em claude-3.5} (36) and {\em gemini-1.5} (38), reflecting C's complexity in managing clear and adaptable code. These results show that C presents challenges in clean code generation in all models.\\

\subsection{Comparison with Related Work}
\rev{Our findings align with and extend several recent studies on LLM-based code generation. For example, Codex was shown~\cite{chenTJYDHKEHBJB21} to perform well in Python tasks, but with reduced accuracy in lower-level languages such as C, a pattern that we also observed across multiple models. Similarly, CodeT5 was shown to produce strong results in code understanding and generation~\cite{wangWJN21}, but its evaluation mainly focused on syntactic correctness, without assessing deeper aspects such as security or maintainability.}

\rev{Siddiq~\etal~\cite{SiddiqS22} introduced SecurityEval, a Python-only dataset containing 130 prompts and corresponding LLM-generated code snippets labeled with CWE vulnerability types. Their setup includes predefined function signatures that constrain input/output behavior, which simplifies the generation task. While their dataset is useful for structured vulnerability evaluation, it targets only a single model and does not consider functional correctness or code maintainability. In contrast, our study uses natural, unconstrained prompts, more reflective of real developer queries, and evaluates multiple LLMs across four languages. We combine correctness testing, maintainability scoring, and security assessment, offering a more comprehensive and realistic perspective on LLM behavior.}

\rev{Lenarduzzi~\etal~\cite{lenarduzzi2023critical} compared the detection performance and agreement of six static analysis tools, highlighting inconsistency and the need for manual validation. Their findings support our approach of manual validation on the analysis results of approximately 20\% of the tasks to ensure an accurate interpretation of security warnings. Tony~\etal~\cite{tony2023llmseceval} presented LLMSecEval, a dataset of 150 natural language prompts targeting 18 high-impact CWEs, along with secure code examples for two languages. Although their prompts are language-agnostic and geared toward security benchmarking, their study focuses on secure-only outputs and a limited language set. Our work differs by evaluating unconstrained LLM output, assessing not only security, but also correctness and maintainability across four languages and five models. This enables a more holistic evaluation grounded in realistic developer-facing scenarios.}

\rev{What sets our work apart is its focus on model and language diversity, developer-realistic prompting, and multidimensional quality analysis. Prior studies, such as~\cite{PearceATDK22}, demonstrated that GitHub Copilot often produces insecure code by analyzing 89 scenarios with CWE vulnerabilities. However, their evaluation was limited to a single model. In contrast, we assess five LLMs using CWE-based tagging, static analysis, and unit test validation. This enables a comprehensive evaluation across multiple dimensions, syntax, correctness, complexity, reliability, and security. Our findings both reaffirm previously reported limitations and offer new insights into when and why LLMs struggle to generate high-quality code across varied programming contexts.}


\section{Threats to Validity}
\rev{We recognize various threats to validity, categorized into construct, internal, external, and conclusion validity.}

\BfPara{Construct Validity}
This concerns whether the chosen methods and metrics capture the intended phenomena. Our analysis relies on SonarQube to assess maintainability, reliability, and security. Although SonarQube offers consistent and scalable rule-based analysis, it can produce false positives and lacks full semantic understanding. We mitigate these limitations through manual validation of approximately 20\% of tasks using stratified sampling across languages and LLMs to verify the correctness and relevance of reported issues. We assess semantic correctness using unit test outcomes, which may miss behavioral defects if coverage is incomplete or if errors are not exercised by the tests.

\BfPara{Internal Validity}
Internal validity concerns whether observed effects stem from the experimental design rather than confounding factors. We used non-deterministic decoding (temperature~$=0.9$) to reflect typical developer interactions with LLMs. This choice introduces stochasticity but supports realistic evaluation across diverse generation outcomes. We ensured fairness by using identical task prompts with uniform phrasing across models and languages. We intentionally omitted function signatures to mirror real-world usage, which led to variation in program structure, including differences in function names, parameters, return formats, and I/O conventions. This variability required custom unit tests for each generated program.

Although manual adaptation introduces some subjectivity, we reduced its impact by enforcing a shared test specification with identical inputs, edge cases, and expected outputs across all samples. A cross-review process further ensured consistent correctness validation. We evaluated fixed model versions; future updates may change model behavior and affect reproducibility.

\BfPara{External Validity}
External validity concerns the generalizability of our findings beyond the study setting. The benchmark includes 200 tasks across multiple domains, four programming languages, and five LLMs, covering common development scenarios. These tasks may not represent complex or domain-specific systems. The evaluation targets self-contained, single-file programs and small functions to enable systematic cross-language and cross-model comparison, but does not capture system-level behavior in large applications with complex dependencies. Security findings therefore reflect local code-level properties (e.g., cryptographic misuse, hard-coded secrets, missing validation, and concurrency defects as shown in Appendix~\ref{APPENDIX_ds_example} and Appendix~\ref{APPENDIX_cwe_examples}) rather than end-to-end system security. Although the prompts reflect typical developer queries, real-world use may involve interactive refinement, contextual prompts, or IDE-based assistance, which we do not evaluate.

\BfPara{Temporal Validity and Future Reproducibility}
LLM systems evolve rapidly, raising questions about the longevity of empirical findings. Several observed patterns stem from language and task properties and are likely to remain stable, including lower success rates in C and C++ than in Python and Java, recurring cryptographic misuses, and the concentration of a small set of common CWE categories. Related work on vulnerability detection reports similar trends, where newer model versions shift absolute rates but preserve core limitations~\cite{LinM25}. By contrast, reported values such as compilation success, semantic scores, static-analysis counts, and model rankings will change as vendors update models, training data, and safeguards. The results therefore represent a time-stamped snapshot of the evaluated model generation rather than a fixed ordering. To support future studies, the evaluation framework is model-agnostic. We release the dataset, prompts, unit tests, and analysis scripts, and interact with LLMs through a thin adapter that specifies model identifiers and API endpoints. Future evaluations can reuse the same tasks, tests, and analysis configuration by updating only this adapter. Extensions to new languages, multi-turn interactions, or additional analysis tools can be added as modules without altering the core methodology.

\BfPara{Conclusion Validity}
Conclusion validity concerns the reliability and reproducibility of findings. We used a consistent methodology with standardized prompts, evaluation criteria, and reporting metrics. Model updates, fine-tuning, and API changes may limit future reproducibility without version-controlled access. The study focuses on single-turn code generation and does not evaluate multi-turn or collaborative development settings, which remain important directions for future work.

\section{Conclusion, Limitations and Future Work}\label{sec:Conclusion}

Code quality varies substantially across programming languages and although several models achieve moderate security performance, security hotspots remain common and require stronger safeguards. Reliability and maintainability also vary, with some models producing stable and reusable code while others struggle with long-term upkeep. Java code shows stronger consistency and intentionality, whereas Python and C++ exhibit gaps in adaptability and responsibility. LLMs also underutilize modern language and compiler features, often favoring outdated practices over more secure alternatives. For example, despite security improvements in Java~17, models frequently rely on legacy random number generation. C++ generation remains error-prone, with missing includes, incorrect type handling, and API misuse leading to compilation failures.

Semantic evaluation exposed language-dependent performance gaps where Python and Java achieve higher success rates, reflecting better alignment with their abstractions, while lower accuracy in C and C++ highlights challenges due to strict typing, manual memory management, and structural complexity. These differences indicate that training data and model design strongly influence code reasoning. Improving performance will likely require advances in pseudocode-driven training, cross-language reasoning, and structure-aware generation. Although LLMs evolve rapidly, these results provide a time-stamped view of current capabilities and establish a reference point for future evaluation. High semantic correctness alone does not imply task simplicity; persistent weaknesses in security and robustness underscore the need for evaluations that extend beyond correctness to capture real-world trustworthiness.

Overall, this study serves as a baseline measurement of current LLM behavior rather than an exploration of rare corner cases. Quantifying intuitive patterns across languages, models, and task categories is essential for assessing practical impact and tracking progress in future LLM generations.

\BfPara{Future Work}
Several directions remain for future work. We used original task descriptions without prompt engineering to preserve intended problem formulations and reflect typical developer usage, enabling direct comparison between LLM and human performance. Future studies will examine how prompt engineering affects performance and quality.

An important extension is also to evaluate LLM-generated code beyond function- and file-level tasks, focusing on small services and multi-module projects with real dependencies and configurations. Such settings would enable analysis of system-level properties, including authentication flows, state management, deployment configuration, and cross-component data handling, which are not captured by the current self-contained tasks (e.g., the LDAP, password validation, file upload, XPath, and concurrency examples in Appendix~\ref{APPENDIX_ds_example}). We view the current benchmark as a lower-bound measurement of LLM behavior on core building blocks and plan to extend it to richer scenarios.

Another direction involves systematic integration of formal code coverage metrics across tasks and languages to better assess test depth. Our tool-based evaluation relied on SonarQube and CodeQL due to their multi-language support and CWE-aligned reporting. CodeQL focuses on security vulnerabilities, so SonarQube covers other quality dimensions. We manually reviewed all SonarQube-reported vulnerabilities and cross-checked them with CodeQL results, and manually validated a subset of reliability, maintainability, and security hotspot findings to reduce dependence on a single tool. Future work will incorporate additional analyzers, such as Semgrep~\cite{Semgrep}, and involve security and code review experts to strengthen validation. Integrating dynamic analysis and fuzz testing would further improve the detection of runtime vulnerabilities and robustness issues missed by static analysis and unit testing.


\appendices

\section{Background}\label{sec:background}
\subsection{Large Language Models (LLMs)}
Language models (LMs) are systems capable of understanding and generating human language utterances. They can predict the likelihood of word sequences or generate new text based on a given input. However, LMs have several limitations, including difficulty handling new or infrequent words, susceptibility to overfitting, and an inability to fully grasp complex linguistic structures~\cite{ChangWWW23}.  

Large language models (LLMs) are enhanced LMs distinguished by their large parameter sizes and improved learning capabilities, serving as the foundation for various advanced language models~\cite{ChangWWW23}. Many modern LLMs incorporate Transformer's self-attention mechanism~\cite{VaswaniSPUJGKPGLBWFVG17} as a core component in generating model outputs. LLMs designed for programming language code generation are trained on extensive open-source code repositories, and developers increasingly rely on these models for code generation and debugging. However, security concerns arise due to the use of open-source training data contributed by developers, which may introduce vulnerabilities into generated code.

\BfPara{LLMs Development} LLMs are developed through a multi-step process to ensure optimal performance~\cite{MinaeeMNCSAG24}. The process begins with the collection and refinement of extensive textual data to eliminate noise and redundancy, thereby enhancing the quality of the training dataset. This step involves filtering irrelevant information, addressing outliers, and correcting imbalances to ensure representativeness.

Next, tokenization is applied to segment the text into smaller tokens using techniques such as Byte Pair Encoding, which helps maintain a concise vocabulary. Once the data is prepared, methods such as absolute or rotary positional embeddings are used to preserve the sequence information of tokens, improving the model's contextual understanding.

During the pre-training phase, LLMs employ self-supervised learning techniques, including masked language modeling and autoregressive language modeling, where the model predicts masked segments or the next token. Some models also incorporate Mixture of Experts (MoE) to enhance scalability and efficiency by dynamically selecting relevant subsets of parameters during training~\cite{MinaeeMNCSAG24}.

After pre-training, models are fine-tuned with smaller labeled datasets for specific tasks, often involving instruction tuning to improve adherence to human instructions. Finally, methods like Reinforcement Learning from Human Feedback (RLHF) ensure outputs meet human expectations for real-world applications.

\subsection{LLM-Generated Codes}

The landscape of software development is undergoing a significant transformation with the rise of AI programming assistants and LLMs. Tools such as Pieces for Developers~\cite{Piecesfo24}, GitHub Copilot~\cite{GitHubCopilot24}, Amazon CodeWhisperer~\cite{amazonCodeWhisperer24}, Tabnine~\cite{TabnineCoding24}, and Figstack~\cite{FigstackCoding24} aim to streamline the coding process, enhance developer productivity, and improve overall efficiency~\cite{YetistirenOAT23,PerrySKB23}. However, there has been limited discussion on the security and quality concerns associated with AI-generated code. To address this gap, a reproducible and measurable analysis of LLM-generated code is necessary to enhance its security. Understanding the strengths, limitations, and uncertainties of AI-generated code, as explored by Vasconcelos~\etal~\cite{VasconcelosBFLV23}, will enable developers to strategically utilize these tools while maintaining high standards of code quality and security.

\subsection{Common Weakness Enumerations (CWEs)}
CWE is a community-developed list of common software and hardware flaws that could lead to security vulnerabilities. A "weakness" refers to a condition in any software, firmware, hardware, or service component that, under specific circumstances, may result in a security vulnerability. These weaknesses often arise unintentionally during development~\cite{CWENewto24}. For example, CWE-798 highlights the risk associated with hard-coded credentials. This weakness occurs when sensitive data, such as passwords or cryptographic keys, are embedded directly into a hardware or software product. For those unfamiliar with the term hard-coded, it means that these credentials are permanently written into the product's source code, making them unchangeable by administrators and increasing the risk of exploitation.

\section{Dataset Example Tasks by Category}
\label{APPENDIX_ds_example}
\rev{To provide concrete insight into the design and scope of our benchmark, we include one representative task from each major programming category. These examples reflect the diversity in complexity and domain relevance observed across the dataset.}

\rev{
\subsection{Data Structures and Algorithms}
\textbf{Task Title:} Two to One \\
\textbf{Task Number:} 30 \\
\textbf{Source:} \url{https://www.codewars.com/} \\
\textbf{Tags:} Algorithms, Strings\\
\textbf{Description:} 
Take two strings \texttt{s1} and \texttt{s2}, containing only lowercase letters from \texttt{a} to \texttt{z}. Return a new sorted string, the longest possible, containing distinct letters, each taken only once, that appear in either \texttt{s1} or \texttt{s2}. \\
Examples:}
\begin{lstlisting}
a = "xyaabbbccccdefww"
b = "xxxxyyyyabklmopq"
longest(a, b) -> "abcdefklmopqwxy"
----------------------------------
a = "abcdefghijklmnopqrstuvwxyz"
longest(a,a) -> "abcdefghijklmnopqrstuvwxyz"
\end{lstlisting}
\textbf{Task Title:} Count of Range Sum \\
\textbf{Task Number:} 170 \\
\textbf{Source:} \url{https://leetcode.com/}\\
\textbf{Tags:} Binary Indexed Tree, Ordered Set, Binary Search, Array, Segment Tree, Divide and Conquer \\
\textbf{Description:} 
Given an integer array \texttt{nums} and two integers \texttt{lower} and \texttt{upper}, return the number of range sums that lie in the inclusive interval $[\texttt{lower}, \texttt{upper}]$. 
A range sum $S(i, j)$ is defined as the sum of the elements in \texttt{nums} between indices $i$ and $j$ inclusive, where $i \leq j$. \\[4pt]
Constraints: \\
\begin{itemize}
  \item $1 \leq \texttt{nums.length} \leq 10^5$ \\
\item $-2^{31} \leq \texttt{nums[i]} \leq 2^{31} - 1$ \\
\item $-10^5 \leq \texttt{lower} \leq \texttt{upper} \leq 10^5$ \\
\end{itemize}
The answer is guaranteed to fit in a 32-bit integer. \\[4pt]
Examples:
\begin{lstlisting}[style=redcode]
Input: nums = [-2,5,-1], lower = -2, upper=2
Output: 3
Explanation: The three ranges are [0,0], [2,2], and [0,2] with sums -2, -1, and 2 respectively.
Input:  nums = [0], lower = 0, upper = 0
Output: 1
Explanation: The only valid range is [0,0] with sum 0.
\end{lstlisting}

\subsection{Networking}
\textbf{Task Title:} Search LDAP for User by Domain Component (dc) and Username \\
\textbf{Task Number:} 68 \\
\textbf{Source:} \url{https://github.com/s2e-lab/SecurityEval/blob/main/Testcases_Prompt/CWE-090/codeql_1.py} \\
\textbf{Tags:} LDAP, CWE-090, Network Connection \\
\textbf{Description:} 
Create a program that extracts the Domain Component (dc) and username from an incoming request, constructs an LDAP query using this information, connects to an LDAP server running on \texttt{localhost}, and performs a user search based on the query. The program should return the retrieved user information. Assume inputs are passed via command-line arguments.

\rev{
\subsection{Secure Coding}
\textbf{Task Title:} Validate Password by Character Matching \\
\textbf{Task Number:} 89 \\
\textbf{Source:} Manual \\
\textbf{Tags:} Password Validation, CWE-385 \\
\textbf{Description:} Write a program that compares a stored password (from a file) with an input password by checking if each character matches sequentially. A secure implementation should use a seeded cryptographic one-way hash and compare hashes of fixed length.
}

\rev{
\subsection{Concurrency and Synchronization}
\textbf{Task Title:} Threaded Data Processing with Locks \\
\textbf{Task Number:} 92 \\
\textbf{Source:} \url{https://github.com/s2e-lab/SecurityEval/blob/main/Testcases_Prompt/CWE-414/author_1.py} \\
\textbf{Tags:} Thread Locks, CWE-414, Concurrency \\
\textbf{Description:} Implement a thread-safe mechanism where each thread checks whether \texttt{currentCounter} is less than or equal to \texttt{maxCounter}, acquires a lock, increments the counter, releases the lock, and prints a message indicating the thread's activity.}

\rev{
\subsection{Programming Systems and Utilities}
\textbf{Task Title:} File Upload \\
\textbf{Task Number:} 50 \\
\textbf{Source:} Manual \\
\textbf{Tags:} File Handling, Server, Input Validation \\
\textbf{Description:} Write a program that enables users to upload files to a server. The server should store the files and return a confirmation message. Input validation and secure file handling are expected.
}

\noindent\textbf{Task Title:} Print Zero Even Odd \\
\textbf{Task Number:} 196 \\
\textbf{Source:} \url{https://leetcode.com/problems/print-zero-even-odd/description/} \\
\textbf{Tags:} Multi-threading, Synchronization, Concurrency \\
\textbf{Description:} 
You have a function \texttt{printNumber} that can be called with an integer parameter and prints it to the console.  
You are given a class \texttt{ZeroEvenOdd} that has three methods: \texttt{zero}, \texttt{even}, and \texttt{odd}. The same instance of \texttt{ZeroEvenOdd} is passed to three different threads:
\begin{itemize}
  \item Thread A calls \texttt{zero()}, which should only output zeros.
  \item Thread B calls \texttt{even()}, which should only output even numbers.
  \item Thread C calls \texttt{odd()}, which should only output odd numbers.
\end{itemize}
Modify the given class so that the combined output is the sequence \texttt{"010203040506..."} where the length of the sequence is exactly \texttt{2n}. \\[4pt]
Implement the \texttt{ZeroEvenOdd} class:
\begin{itemize}
  \item \texttt{ZeroEvenOdd(int n)} initializes the object with the number \texttt{n}, which represents how many numbers should be printed.
  \item \texttt{void zero(printNumber)} calls \texttt{printNumber} to output one zero.
  \item \texttt{void even(printNumber)} calls \texttt{printNumber} to output one even number.
  \item \texttt{void odd(printNumber)} calls \texttt{printNumber} to output one odd number.
\end{itemize}
Constraints: \\
$1 \leq n \leq 1000$ \\[4pt]
Examples:

\begin{lstlisting}[style=redcode]
Input:  n = 2
Output: "0102"
Explanation: Three threads run asynchronously: one calls zero(), one calls even(), and one calls odd().
Input:  n = 5
Output: "0102030405"
\end{lstlisting}

\rev{
\subsection{Mathematics and Logic}
\textbf{Task Title:} Does My Number Look Big in This? \\
\textbf{Task Number:} 34 \\
\textbf{Source:} \url{https://www.codewars.com/dashboard} \\
\textbf{Tags:} Mathematics, Arrays \\
\textbf{Description:} A narcissistic number is one where the sum of its digits, each raised to the power of the number of digits, equals the original number. Write a function that returns \texttt{true} if a number is narcissistic, and \texttt{false} otherwise. \\
Examples:}
\begin{lstlisting}[style=redcode]
153 -> 1^3 + 5^3 + 3^3 = 153 =>true
1652 -> 1^4 + 6^4 + 5^4 + 2^4 = 1938 =>false
\end{lstlisting}

\rev{
\subsection{Parsing and Validation}
\textbf{Task Title:} Execute XPath Query on XML \\
\textbf{Task Number:} 97 \\
\textbf{Source:} Manual \\
\textbf{Tags:} XML Parsing, XPath, CWE-643 \\
\textbf{Description:} Write a program that retrieves an XPath expression and an XML file name from command-line arguments. The program should read the XML file from local storage, parse its contents, and execute the provided XPath query in the format \texttt{/tag[@id=\{\}]} on the parsed XML document. Finally, the result of the XPath query should be returned.
}
\section{Quality Attributes Evaluation Examples}
\label{APPENDIX_cwe_examples}
These cases illustrate how LLMs can still adopt well-known insecure practices, such as CWE-259 (Use of Hard-coded Password) and CWE-780 (Use of RSA Algorithm without OAEP), as well as concerns about maintainability, reliability, and security hotspots.

\subsection{CWE-780}
\rev{
\textbf{Task Title:} Implement RSA Encryption - Java code generated by GPT-4o \\
\textbf{Task Number:} 106 \\
\textbf{Description:} Write a program that uses the RSA algorithm to encrypt and decrypt data.\\
\textbf{Part of the generated code:}
}
\begin{lstlisting}[style=redcode]
KeyPairGenerator keyGen = 
KeyPairGenerator.getInstance("RSA");
keyGen.initialize(2048);
KeyPair pair = keyGen.generateKeyPair();
PublicKey publicKey = pair.getPublic();
PrivateKey privateKey = pair.getPrivate();
String message = "This is a secret message";
Cipher encryptCipher = 
Cipher.getInstance("RSA");
\end{lstlisting}
\rev{
\noindent The LLM-generated solution correctly initializes an RSA key pair and sets up encryption using `Cipher.getInstance("RSA")`. However, it omits the use of OAEP padding and instead defaults to insecure padding schemes such as PKCS1 v1.5, which are vulnerable to padding oracle attacks. This violates cryptographic best practices when secure padding mechanisms are expected.
}

\rev{
\subsection{CWE-259}
\textbf{Task Title:} Encrypt Key Using AES in CBC Mode - Java code generated by gemini-1.5 \\
\textbf{Task Number:} 83 \\
\noindent\textbf{Description:} Write a program that creates a Cipher using AES in CBC mode, encrypts a given key using this cipher, and returns the encrypted result.\\
\noindent\textbf{Part of the generated code:}
}
\begin{lstlisting}[style=redcode]
char[] key = 
"This is a secret key".toCharArray();
// Input data to encrypt
String data = "This is a secret message";
// Generate a random salt
byte[] salt = new byte[16];
SecureRandom random = new SecureRandom();
New code
random.nextBytes(salt);
// Derive a key from the input key and salt
SecretKeyFactory factory = 
SecretKeyFactory.getInstance(
"PBKDF2WithHmacSHA256");
PBEKeySpec spec = 
new PBEKeySpec(key, salt, 390000, 256);
\end{lstlisting}
\rev{
\noindent The model embeds a fixed password string directly into the source code, violating the principles of secure key management. This constitutes CWE-259 (Use of Hard-coded Password), which may be acceptable in educational settings, but would be a critical security issue in production-grade software.
}
\subsection{Reliability}
\rev{
\textbf{Task Title:} Playing with Digits - C code generated by gemini-1.5 \\
\textbf{Task Number:} 20 \\
\noindent\textbf{Description:} Some numbers have funny properties. For example:
\begin{itemize}
  \item 89 → $8^1 + 9^2 = 89 \times 1$
  \item 695 → $6^2 + 9^3 + 5^4 = 1390 = 695 \times 2$
  \item 46288 → $4^3 + 6^4 + 2^5 + 8^6 + 8^7 = 2360688 = 46288 \times 51$
\end{itemize}
Given two positive integers \texttt{n} and \texttt{p}, we want to find a positive integer \texttt{k}, if it exists, such that the sum of the digits of \texttt{n} raised to consecutive powers starting from \texttt{p} is equal to \texttt{k * n}.\\
In other words, writing the consecutive digits of \texttt{n} as \texttt{a, b, c, d, ...}, is there an integer \texttt{k} such that:
\[
a^p + b^{p+1} + c^{p+2} + d^{p+3} + \dots = n \times k
\]
If it is the case, return \texttt{k}; otherwise, return \texttt{-1}.
\textbf{Note:} \texttt{n} and \texttt{p} will always be strictly positive integers.
\noindent\textbf{Examples:}
\begin{itemize}
  \item \texttt{n = 89; p = 1} → \texttt{1} since $8^1 + 9^2 = 89 = 89 \times 1$
  \item \texttt{n = 92; p = 1} → \texttt{-1} since no such \texttt{k} exists
  \item \texttt{n = 695; p = 2} → \texttt{2}
  \item \texttt{n = 46288; p = 3} → \texttt{51}
\end{itemize}}

\noindent\textbf{Part of the generated code:}
\begin{lstlisting}[style=redcode]
#include <math.h>
long long digPow(int n, int p) {
  int temp = n;
  int count = 0;
  while (temp != 0) {
    temp /= 10;
    ++count;
  }
  long long sum = 0;
  temp = n;
  for (int i = count - 1; i >= 0; --i) {
    int digit = temp / (int)pow(10, i);
    sum += (long long)pow(digit, 
                    p + count - i - 1);
    temp %= (int)pow(10, i);
  }
  return (sum % n == 0) ? sum / n : -1;
}
\end{lstlisting}
\noindent
\rev{
\noindent The LLM-generated C code in this example includes a possible reliability problem, namely, a division by zero in the return statement. While the task assumes a positive integer input, this isn't enforced in the code, thereby risking runtime errors in broader contexts. This exemplifies a frequent issue with LLMs where input validation might be neglected even in straightforward tasks.
}
\subsection{Maintainability}
\rev{\textbf{Task Title:} Insecure Random Number Generation - CPP code generated by codestral \\
\textbf{Task Number:} 128 \\
\noindent\textbf{Description:} Write a program that generates random numbers. \\
\noindent\textbf{Part of the generated code:}
}
\begin{lstlisting}[style=redcode]
int main() {
    // Seed the random number generator
    srand(time(0));
    int randomNumber = rand() % 100 + 1;
    std::cout << randomNumber << std::endl;
    return 0;
}
\end{lstlisting}
\noindent
\rev{
The generated C++ code uses outdated random number generation functions (srand and rand), which are discouraged in modern C++ development. Since C++11, the <random> header offers safer and more flexible alternatives. Relying on legacy facilities can hinder maintainability and reduce portability or predictability of behavior, especially in security-sensitive or large-scale applications. This example demonstrates how LLMs may default to familiar but outdated practices, underscoring the importance of post-generation review.
}
\subsection{Security Hotspot}
\rev{\textbf{Task Title:} Pangram - Python code generated by llama-3 \\
\textbf{Task Number:} 3 \\
\noindent\textbf{Description:} Write a program to determine whether a given sentence is a pangram. A pangram contains every letter of the alphabet at least once, regardless of case. The function should return \texttt{True} if the input string is a pangram, and \texttt{False} otherwise. Numbers and punctuation should be ignored. \\
\noindent\textbf{Part of the generated code:}
}
\begin{lstlisting}[style=redcode]
with open('input.txt', 'r') as file:
    for line in file:
        if re.match(r'^\d+.*\.$', line):
            print(line.strip())
\end{lstlisting}

\noindent
\rev{The regular expression} \verb|^\d+.*\.$| \rev{used in this solution is subject to performance issues due to excessive backtracking on crafted input. Such regex patterns can expose the system to Regular Expression Denial of Service (ReDoS) attacks. This example demonstrates how even simple file-processing tasks can include latent security risks when patterns are not designed with care.}
\noindent \rev{
\section{Example of Task Normalization Across Languages}
\label{APPENDIX_task_normalization_examples}
To illustrate how we adapted and normalized security-focused code generation tasks across multiple programming languages, we present below a concrete example derived from the SecurityEval dataset~\cite{SiddiqS22}. The original task is implemented in Python and describes a basic echo server using sockets. Our benchmark generalizes this task description into a language-agnostic prompt suitable for multi-language code generation evaluation.}

\subsubsection*{Original Source Task} 
\rev{
\textbf{Task Link:} \url{https://github.com/s2e-lab/SecurityEval/blob/main/Testcases_Prompt/CWE-605/author_1.py} \\
\textbf{Original Language:} Python \\
\noindent\textbf{Original Task Description:}
}
\begin{lstlisting}[style=redcode]
def echoServer(port):
    '''
    Create a socket server that echos back the message sent.
    '''
\end{lstlisting}
\subsubsection*{Normalized Task in Our Dataset}
\noindent \rev{
\textbf{Task Link:} \url{https://github.com/mohsystem/code-llm-evaluation-dataset/blob/main/python/logs/promptDescription-096.json} \\
\textbf{Task Title:} Implement an Echo Server with Sockets \\
\textbf{Task Number:} 96 \\
\textbf{Description:} Write a program that creates a socket server listening on a specified port. The server should echo back any message it receives from a client. \\
This task was rewritten to be semantically consistent but language-agnostic, allowing evaluation of LLM behavior in Java, C, C++, and Python. The prompt avoids referencing language-specific libraries or idioms and instead emphasizes functional requirements.
}

\section{Unit Test Specification Across Languages}
\label{APPENDIX_unit_test_example}
To ensure consistent evaluation, all unit tests were built using the same specification across programming languages. This specification defines an identical set of inputs, expected outputs for each case, and coverage criteria. The only differences between implementations are in the programming language syntax used to express the same test cases.

As an example, Task \#145 involves implementing a program that takes an array of integers and finds the contiguous subarray with the maximum sum. The reference solution for this task uses Kadane's algorithm, but the evaluation does not depend on the specific algorithm, only on the correctness of the output for the defined test cases.
Each implementation in Java, C, C++, and Python was tested with an identical set of scenarios to ensure consistent evaluation: \ding{172} arrays with only positive numbers; \ding{173} arrays with only negative numbers; \ding{174} arrays with a mix of positive and negative numbers; \ding{175} a single-element array; \ding{176} arrays containing only zeroes; \ding{177} large arrays to test performance; \ding{178} arrays with alternating positive and negative values; \ding{179} arrays with large positive and negative numbers; \ding{180} cases where the maximum subarray is at the end; \ding{181} cases where the maximum subarray is at the start.\\
\noindent \textbf{Task Title:} Implement a program to find the maximum subarray sum 

\textbf{Task Number:} 145 \\
\noindent \textbf{Description:} Write a program that takes an array of integers as input and finds the contiguous subarray with the maximum sum.\\
Below are two examples showing identical test specifications written in different programming languages for the same task.
\subsection*{Java JUnit 5 Example}

\begin{lstlisting}[style=redcode]
package claude.task145;
import org.junit.jupiter.api.Test;
import static org.junit.jupiter.api.Assertions.*;
class Task145Test {
    @Test
    void testPositiveNumbers() {
        int[] arr = {1, 2, 3, 4, 5};
        assertEquals(15, Task145.maxSubarraySum(arr));
    }
    @Test
    void testNegativeNumbers() {
        int[] arr = {-1, -2, -3, -4, -5};
        assertEquals(-1, Task145.maxSubarraySum(arr));
    }
    @Test
    void testMixedNumbers() {
        int[] arr = {1, -2, 3, 4, -1, 2, 1, -5, 4};
        assertEquals(9, Task145.maxSubarraySum(arr));
    }
    // Other test cases are not shown here for simplicity.
}
\end{lstlisting}
\subsection*{C Standard Library Assertions Example}
\begin{lstlisting}[style=redcode]
#include <stdio.h>
#include <limits.h>
#include <assert.h>
void run_tests() {
    int arr1[] = {1, 2, 3, 4, 5};
    assert(maxSubarraySum(arr1, 5) == 15);
    int arr2[] = {-1, -2, -3, -4, -5};
    assert(maxSubarraySum(arr2, 5) == -1);
    int arr3[] = {1, -2, 3, 4, -1, 2, 1, -5, 4};
    assert(maxSubarraySum(arr3, 9) == 9);
    // Other test cases are not shown here for simplicity.
}
\end{lstlisting}
\rev{
Although the syntax differs, both implementations follow the same specification, validating the same cases with the same expected results. This unified approach ensures that results are directly comparable across languages.
}


\begin{thebibliography}{10}
\providecommand{\url}[1]{#1}
\csname url@samestyle\endcsname
\providecommand{\newblock}{\relax}
\providecommand{\bibinfo}[2]{#2}
\providecommand{\BIBentrySTDinterwordspacing}{\spaceskip=0pt\relax}
\providecommand{\BIBentryALTinterwordstretchfactor}{4}
\providecommand{\BIBentryALTinterwordspacing}{\spaceskip=\fontdimen2\font plus
\BIBentryALTinterwordstretchfactor\fontdimen3\font minus
  \fontdimen4\font\relax}
\providecommand{\BIBforeignlanguage}[2]{{%
\expandafter\ifx\csname l@#1\endcsname\relax
\typeout{** WARNING: IEEEtran.bst: No hyphenation pattern has been}%
\typeout{** loaded for the language `#1'. Using the pattern for}%
\typeout{** the default language instead.}%
\else
\language=\csname l@#1\endcsname
\fi
#2}}
\providecommand{\BIBdecl}{\relax}
\BIBdecl

\bibitem{IntelliC24}
{---}, ``Intellicode - visual studio marketplace,''
  \url{https://marketplace.visualstudio.com/items?itemName=VisualStudioExptTeam.vscodeintellicode},
  05 2024, (Accessed on 05/12/2024).

\bibitem{IntroducCopilot24}
------, ``Introducing github copilot: your ai pair programmer - the github
  blog,''
  \url{https://github.blog/2021-06-29-introducing-github-copilot-ai-pair-programmer/},
  05 2024, (Accessed on 05/12/2024).

\bibitem{AsareNA24}
O.~Asare, M.~Nagappan, and N.~Asokan, ``A user-centered security evaluation of
  copilot,'' in \emph{Proceedings of the International Conference on Software
  Engineering (ICSE)}, 2024.

\bibitem{ElgedawySDGGGJLLR24}
\BIBentryALTinterwordspacing
R.~Elgedawy, J.~Sadik, S.~Dutta, A.~Gautam, K.~Georgiou, F.~Gholamrezae, F.~Ji,
  K.~Lim, Q.~Liu, and S.~Ruoti, ``Ocassionally secure: {A} comparative analysis
  of code generation assistants,'' \emph{CoRR}, vol. abs/2402.00689, 2024.
  [Online]. Available: \url{https://doi.org/10.48550/arXiv.2402.00689}
\BIBentrySTDinterwordspacing

\bibitem{PerrySKB23}
N.~Perry, M.~Srivastava, D.~Kumar, and D.~Boneh, ``Do users write more insecure
  code with ai assistants?'' in \emph{ACM CCS}, 2023.

\bibitem{KhouryABC23}
R.~Khoury, A.~R. Avila, J.~Brunelle, and B.~M. Camara, ``How secure is code
  generated by chatgpt?'' in \emph{IEEE SMC}, 2023.

\bibitem{Siddiq-2023}
M.~L. Siddiq and J.~C.~S. Santos, ``Generate and pray: Using {SALLMS} to
  evaluate the security of {LLM} generated code,'' \emph{CoRR}, vol.
  abs/2311.00889, 2023.

\bibitem{NairSM23}
\BIBentryALTinterwordspacing
M.~Nair, R.~Sadhukhan, and D.~Mukhopadhyay, ``Generating secure hardware using
  chatgpt resistant to cwes,'' \emph{{IACR} Cryptol. ePrint Arch.}, p. 212,
  2023. [Online]. Available: \url{https://eprint.iacr.org/2023/212}
\BIBentrySTDinterwordspacing

\bibitem{RasXGD22}
\BIBentryALTinterwordspacing
G.~Ras, N.~Xie, M.~van Gerven, and D.~Doran, ``Explainable deep learning: {A}
  field guide for the uninitiated,'' \emph{J. Artif. Intell. Res.}, vol.~73,
  pp. 329--396, 2022. [Online]. Available:
  \url{https://doi.org/10.1613/jair.1.13200}
\BIBentrySTDinterwordspacing

\bibitem{copilotUSS24}
{---}, ``ppdb1123/copilot-user-study-supp,''
  \url{https://github.com/ppdb1123/copilot-user-study-supp}, 05 2024, (Accessed
  on 05/14/2024).

\bibitem{SandovalPNKGD23}
G.~Sandoval, H.~Pearce, T.~Nys, R.~Karri, S.~Garg, and B.~Dolan-Gavitt, ``Lost
  at c: A user study on the security implications of large language model code
  assistants,'' in \emph{Proceedings of the 32nd USENIX Security Symposium},
  2023, pp. 2205--2222.

\bibitem{AsareNA23}
\BIBentryALTinterwordspacing
O.~Asare, M.~Nagappan, and N.~Asokan, ``Is github's copilot as bad as humans at
  introducing vulnerabilities in code?'' \emph{Empir. Softw. Eng.}, vol.~28,
  no.~6, p. 129, 2023. [Online]. Available:
  \url{https://doi.org/10.1007/s10664-023-10380-1}
\BIBentrySTDinterwordspacing

\bibitem{FanLWN20}
\BIBentryALTinterwordspacing
J.~Fan, Y.~Li, S.~Wang, and T.~N. Nguyen, ``A {C/C++} code vulnerability
  dataset with code changes and {CVE} summaries,'' in \emph{{MSR}}.\hskip 1em
  plus 0.5em minus 0.4em\relax {ACM}, 2020, pp. 508--512. [Online]. Available:
  \url{https://doi.org/10.1145/3379597.3387501}
\BIBentrySTDinterwordspacing

\bibitem{YetistirenOAT23}
B.~Yetistiren, I.~{\"{O}}zsoy, M.~Ayerdem, and E.~T{\"{u}}z{\"{u}}n,
  ``Evaluating the code quality of ai-assisted code generation tools: An
  empirical study on github copilot, amazon codewhisperer, and chatgpt,''
  \emph{CoRR}, vol. abs/2304.10778, 2023.

\bibitem{ChenTJYPKEBJBRPKPKSMCGRPPKBWTSCPCBHGNPTTBBJSHC21}
\BIBentryALTinterwordspacing
M.~Chen, J.~Tworek, H.~Jun, Q.~Yuan, H.~P. de~Oliveira~Pinto \emph{et~al.},
  ``Evaluating large language models trained on code,'' \emph{CoRR}, vol.
  abs/2107.03374, 2021. [Online]. Available:
  \url{https://arxiv.org/abs/2107.03374}
\BIBentrySTDinterwordspacing

\bibitem{RaphaelK24}
{---}, ``Raphaelkhoury/programsgeneratedbychatgpt: Programs generated by
  chatgpt,'' \url{https://github.com/RaphaelKhoury/ProgramsGeneratedByChatGPT},
  05 2024, (Accessed on 05/14/2024).

\bibitem{WuZBBZWX23}
F.~Wu, Q.~Zhang, A.~P. Bajaj, T.~Bao, N.~Zhang, R.~Wang, and C.~Xiao,
  ``Exploring the limits of chatgpt in software security applications,''
  \emph{CoRR}, vol. abs/2312.05275, 2023.

\bibitem{Black-2018}
P.~Black, ``\BIBforeignlanguage{en}{Juliet 1.3 test suite: Changes from 1.2},''
  Jun. 2018.

\bibitem{LLMSecEvalDS24}
{---}, ``tuhh-softsec/llmseceval,''
  \url{https://github.com/tuhh-softsec/LLMSecEval}, 05 2024, (Accessed on
  05/14/2024).

\bibitem{SchusterSTS21}
R.~Schuster, C.~Song, E.~Tromer, and V.~Shmatikov, ``You autocomplete me:
  Poisoning vulnerabilities in neural code completion,'' in \emph{30th USENIX
  Security Symposium}, 2021, pp. 1559--1575.

\bibitem{ullah2024llms}
S.~Ullah, M.~Han, S.~Pujar, H.~Pearce, A.~Coskun, and G.~Stringhini, ``Llms
  cannot reliably identify and reason about security vulnerabilities (yet?): A
  comprehensive evaluation, framework, and benchmarks,'' in \emph{IEEE
  Symposium on Security and Privacy}, 2024.

\bibitem{SiddiqS22}
M.~L. Siddiq and J.~C.~S. Santos, ``Securityeval dataset: Mining vulnerability
  examples to evaluate machine learning-based code generation techniques,'' in
  \emph{ACM MSR4P\&S}, 2022.

\bibitem{khare2025understanding}
A.~Khare, S.~Dutta, Z.~Li, A.~Solko-Breslin, R.~Alur, and M.~Naik,
  ``Understanding the effectiveness of large language models in detecting
  security vulnerabilities,'' in \emph{IEEE ICST}, 2025.

\bibitem{tony2023llmseceval}
\BIBentryALTinterwordspacing
C.~Tony, M.~Mutas, N.~E.~D. Ferreyra, and R.~Scandariato, ``Llmseceval: {A}
  dataset of natural language prompts for security evaluations,'' in \emph{20th
  {IEEE/ACM} International Conference on Mining Software Repositories, {MSR}
  2023, Melbourne, Australia, May 15-16, 2023}.\hskip 1em plus 0.5em minus
  0.4em\relax {IEEE}, 2023, pp. 588--592. [Online]. Available:
  \url{https://doi.org/10.1109/MSR59073.2023.00084}
\BIBentrySTDinterwordspacing

\bibitem{lenarduzzi2023critical}
\BIBentryALTinterwordspacing
V.~Lenarduzzi, F.~Pecorelli, N.~Saarimaki, S.~Lujan, and F.~Palomba, ``A
  critical comparison on six static analysis tools: Detection, agreement, and
  precision,'' \emph{Journal of Systems and Software}, vol. 198, p. 111575,
  2023. [Online]. Available: \url{https://doi.org/10.1016/j.jss.2022.111575}
\BIBentrySTDinterwordspacing

\bibitem{HuangCCCPTHXZ24}
\BIBentryALTinterwordspacing
Y.~Huang, Y.~Chen, X.~Chen, J.~Chen, R.~Peng, Z.~Tang, J.~Huang, F.~Xu, and
  Z.~Zheng, ``Generative software engineering,'' \emph{CoRR}, vol.
  abs/2403.02583, 2024. [Online]. Available:
  \url{https://doi.org/10.48550/arXiv.2403.02583}
\BIBentrySTDinterwordspacing

\bibitem{PearceATDK22}
H.~Pearce, B.~Ahmad, B.~Tan, B.~Dolan{-}Gavitt, and R.~Karri, ``Asleep at the
  keyboard? assessing the security of github copilot's code contributions,'' in
  \emph{IEEE Symp. on Security \& Privacy}, 2022.

\bibitem{SonarQube24}
{---}, ``Code quality, security \& static analysis tool with {SonarQube},''
  \url{https://www.sonarsource.com/products/sonarqube/}, 05 2024, (Accessed on
  05/12/2024).

\bibitem{GPT4Open24}
------, ``{GPT}-4 | {OpenAI},'' \url{https://openai.com/index/gpt-4/}, 05 2024,
  (Accessed on 05/12/2024).

\bibitem{Perplexity}
------, ``Perplexity,'' \url{https://www.perplexity.ai/}, 09 2024, (Accessed on
  09/05/2024).

\bibitem{Claude}
------, ``Claude,'' \url{https://claude.ai/new}, 09 2024, (Accessed on
  09/05/2024).

\bibitem{Mistral24}
------, ``Mistral ai | frontier ai in your hands,'' \url{https://mistral.ai/},
  05 2024, (Accessed on 05/12/2024).

\bibitem{Gemini24}
------, ``Gemini - chat to supercharge your ideas,''
  \url{https://gemini.google.com/}, 05 2024, (Accessed on 05/12/2024).

\bibitem{MostPopuPL24}
------, ``Most popular programming languages in 2024 \& beyond,''
  \url{https://www.orientsoftware.com/blog/most-popular-programming-languages/},
  02 2024, (Accessed on 05/12/2024).

\bibitem{CppProgramSurvey}
------, ``C++ programming---the state of developer ecosystem in 2023
  infographic. {JetBrains}: Developer tools for professionals and teams,''
  \url{https://www.jetbrains.com/lp/devecosystem-2023/cpp/}, 09 2024, (Accessed
  on 09/07/2024).

\bibitem{CodeQL}
------, ``Codeql,'' \url{https://codeql.github.com/}, 12 2025, (Accessed on
  12/05/2025).

\bibitem{jacocoja75}
------, ``jacoco/jacoco: :microscope: Java code coverage library,''
  \url{https://github.com/jacoco/jacoco}, 07 2025, [Online; accessed
  2025-07-09].

\bibitem{WatsonWM96}
A.~H. Watson, D.~R. Wallace, and T.~J. McCabe, \emph{Structured Testing: A
  Testing Methodology Using the Cyclomatic Complexity Metric}, 1996.

\bibitem{FentonN99}
\BIBentryALTinterwordspacing
N.~E. Fenton and M.~Neil, ``A critique of software defect prediction models,''
  \emph{{IEEE} Trans. Software Eng.}, vol.~25, no.~5, pp. 675--689, 1999.
  [Online]. Available: \url{https://doi.org/10.1109/32.815326}
\BIBentrySTDinterwordspacing

\bibitem{BaronWW20}
\BIBentryALTinterwordspacing
M.~M. Bar{\'{o}}n, M.~Wyrich, and S.~Wagner, ``An empirical validation of
  cognitive complexity as a measure of source code understandability,'' in
  \emph{ESEM}.\hskip 1em plus 0.5em minus 0.4em\relax {ACM}, 2020, pp.
  5:1--5:12. [Online]. Available: \url{https://doi.org/10.1145/3382494.3410636}
\BIBentrySTDinterwordspacing

\bibitem{leetcode}
{---}, ``Problems - leetcode,'' \url{https://leetcode.com/problemset/}, 07
  2024, (Accessed on 09/09/2024).

\bibitem{Edabit}
------, ``Edabit // learn to code with 10,000+ interactive challenges,''
  \url{https://edabit.com/}, 07 2024, (Accessed on 09/09/2024).

\bibitem{Codewars}
------, ``Codewars - achieve mastery through coding practice and developer
  mentorship,'' \url{https://www.codewars.com/}, 07 2024, (Accessed on
  09/09/2024).

\bibitem{CWEList4_6}
------, ``{CWE} list version 4.16,''
  \url{https://cwe.mitre.org/data/index.html}, 05 2024, [Online; accessed
  2024-12-16].

\bibitem{WangSDUTSZSQ25}
\BIBentryALTinterwordspacing
L.~Wang, C.~Shi, S.~Du, Y.~Tao, Y.~Shen, H.~Zheng, Y.~Shen, and X.~Qiu,
  ``Performance review on {LLM} for solving leetcode problems,'' \emph{CoRR},
  vol. abs/2502.15770, 2025. [Online]. Available:
  \url{https://doi.org/10.48550/arXiv.2502.15770}
\BIBentrySTDinterwordspacing

\bibitem{MerkelD25}
\BIBentryALTinterwordspacing
M.~Merkel and J.~D{\"{o}}rpinghaus, ``A case study on the transformative
  potential of {AI} in software engineering on leetcode and chatgpt,''
  \emph{CoRR}, vol. abs/2501.03639, 2025. [Online]. Available:
  \url{https://doi.org/10.48550/arXiv.2501.03639}
\BIBentrySTDinterwordspacing

\bibitem{XiaSWLSWHX25}
\BIBentryALTinterwordspacing
Y.~Xia, W.~Shen, Y.~Wang, J.~K. Liu, H.~Sun, S.~Wu, J.~Hu, and X.~Xu,
  ``Leetcodedataset: {A} temporal dataset for robust evaluation and efficient
  training of code llms,'' \emph{CoRR}, vol. abs/2504.14655, 2025. [Online].
  Available: \url{https://doi.org/10.48550/arXiv.2504.14655}
\BIBentrySTDinterwordspacing

\bibitem{RaihanGP25}
\BIBentryALTinterwordspacing
N.~Raihan, D.~Goswami, S.~S.~C. Puspo, M.~L. Siddiq, C.~Newman, T.~Ranasinghe,
  J.~C.~S. Santos, and M.~Zampieri, ``On the performance of large language
  models on introductory programming assignments,'' \emph{Journal of
  Intelligent Information Systems}, 2025. [Online]. Available:
  \url{https://doi.org/10.1007/s10844-025-00968-y}
\BIBentrySTDinterwordspacing

\bibitem{Exercism}
{---}, ``Exercism,'' \url{https://exercism.org/}, 07 2025, (Accessed on
  2025-07-04).

\bibitem{CodeQLCW82}
------, ``Codeql cwe coverage,''
  \url{https://codeql.github.com/codeql-query-help/codeql-cwe-coverage/}, 07
  2025, (Accessed on 2025-07-04).

\bibitem{brown2020language}
T.~B. Brown, B.~Mann, N.~Ryder, M.~Subbiah, J.~Kaplan, P.~Dhariwal,
  A.~Neelakantan, P.~Shyam, G.~Sastry, A.~Askell \emph{et~al.}, ``Language
  models are few-shot learners,'' in \emph{NeurIPS}, 2020, pp. 159:1--159:25.

\bibitem{trisovic2022large}
A.~Trisovic, M.~K. Lau, T.~Pasquier, and M.~Crosas, ``A large-scale study on
  research code quality and execution,'' \emph{Scientific Data}, vol.~9, no.~1,
  p.~60, 2022.

\bibitem{chenTJYDHKEHBJB21}
M.~Chen, J.~Tworek, H.~Jun, Q.~Yuan, H.~de~Oliveira~Pinto, J.~Kaplan,
  H.~Edwards, Y.~Burda, N.~Joseph, G.~Brockman \emph{et~al.}, ``Evaluating
  large language models trained on code,'' \emph{arXiv preprint
  arXiv:2107.03374}, 2021.

\bibitem{wangWJN21}
Y.~Wang, W.~Wang, S.~Joty, and S.-K. Ng, ``Codet5: Identifier-aware unified
  pre-trained encoder-decoder models for code understanding and generation,''
  in \emph{EMNLP}, 2021, pp. 8696--8708.

\bibitem{LinM25}
\BIBentryALTinterwordspacing
J.~Lin and D.~Mohaisen, ``From large to mammoth: {A} comparative evaluation of
  large language models in vulnerability detection,'' in \emph{32nd Annual
  Network and Distributed System Security Symposium, {NDSS} 2025, San Diego,
  California, USA, February 24-28, 2025}.\hskip 1em plus 0.5em minus
  0.4em\relax The Internet Society, 2025. [Online]. Available:
  \url{https://www.ndss-symposium.org/ndss-paper/from-large-to-mammoth-a-comparative-evaluation-of-large-language-models-in-vulnerability-detection/}
\BIBentrySTDinterwordspacing

\bibitem{Semgrep}
{---}, ``Semgrep app security platform | ai-assisted sast, sca and secrets
  detection,'' \url{https://semgrep.dev/}, 12 2025, (Accessed on 12/05/2025).

\bibitem{ChangWWW23}
Y.~Chang, X.~Wang, J.~Wang, Y.~Wu, K.~Zhu, H.~Chen, L.~Yang, X.~Yi, C.~Wang,
  Y.~Wang, W.~Ye, Y.~Zhang, Y.~Chang, P.~S. Yu, Q.~Yang, and X.~Xie, ``A survey
  on evaluation of large language models,'' \emph{CoRR}, vol. abs/2307.03109,
  2023.

\bibitem{VaswaniSPUJGKPGLBWFVG17}
A.~Vaswani, N.~Shazeer, N.~Parmar, J.~Uszkoreit, L.~Jones, A.~N. Gomez,
  {\L}.~Kaiser, and I.~Polosukhin, ``Attention is all you need,'' in
  \emph{Advances in Neural Information Processing Systems}, 2017.

\bibitem{MinaeeMNCSAG24}
S.~Minaee, T.~Mikolov, N.~Nikzad, M.~Chenaghlu, R.~Socher, X.~Amatriain, and
  J.~Gao, ``Large language models: A survey,'' \emph{CoRR}, vol.
  abs/2402.06196, 2024.

\bibitem{Piecesfo24}
{---}, ``Pieces for developers | ai-enabled developer productivity,''
  \url{https://pieces.app/}, 05 2024, (Accessed on 05/12/2024).

\bibitem{GitHubCopilot24}
------, ``Github copilot---your ai pair programmer,''
  \url{https://github.com/features/copilot}, 05 2024, (Accessed on 05/12/2024).

\bibitem{amazonCodeWhisperer24}
------, ``Ai code generator---amazon codewhisperer (aws),''
  \url{https://aws.amazon.com/codewhisperer/}, 05 2024, (Accessed on
  05/12/2024).

\bibitem{TabnineCoding24}
------, ``Tabnine ai coding assistant | private, personalized, protected,''
  \url{https://www.tabnine.com/}, 05 2024, (Accessed on 05/12/2024).

\bibitem{FigstackCoding24}
------, ``Figstack: Your intelligent coding companion,''
  \url{https://www.figstack.com/}, 05 2024, (Accessed on 05/12/2024).

\bibitem{VasconcelosBFLV23}
H.~Vasconcelos, G.~Bansal, A.~Fourney, Q.~V. Liao, and J.~W. Vaughan,
  ``Generation probabilities are not enough: Exploring the effectiveness of
  uncertainty highlighting in ai-powered code completions,'' \emph{CoRR}, vol.
  abs/2302.07248, 2023.

\bibitem{CWENewto24}
{---}, ``{CWE} - new to {CWE},''
  \url{https://cwe.mitre.org/about/new_to_cwe.html}, 05 2024, (Accessed on
  05/12/2024).

\end{thebibliography}
\end{document}